\RequirePackage{fix-cm}

\documentclass[12pt, a4paper, authoryear]{elsarticle}
\usepackage[titletoc]{appendix}

\usepackage{etoolbox}
\patchcmd{\emailauthor}{(#2)}{}{}{}

\usepackage[dvipsnames]{xcolor}
\usepackage{graphicx}
\usepackage{amsmath}
\usepackage{amsfonts}
\usepackage{amsthm}
\usepackage{mathtools}
\usepackage{upgreek}
\usepackage{leftidx}
\usepackage{amssymb}
\usepackage{scalerel}
\usepackage{lscape} 

\usepackage{tikz,graphicx}
\usetikzlibrary{arrows, automata}

\usepackage{booktabs}
\usepackage{multirow}
\usepackage{tabularx}
\usepackage{pifont}
\usepackage{geometry}
\usepackage{moresize}
\usepackage[hang,flushmargin]{footmisc}
\usepackage{caption}
\usepackage{soul}
\usepackage{paralist}
\usepackage{setspace}
\usepackage{subcaption}
\usepackage{epstopdf}
\usepackage{floatrow}
\usepackage{hyperref}

\usepackage{pgfplots}
\pgfplotsset{compat=newest}
\usepgfplotslibrary{groupplots}
\usepgfplotslibrary{polar}
\usepgfplotslibrary{smithchart}
\usepgfplotslibrary{statistics}
\usepgfplotslibrary{dateplot}

\usepackage[X2,T1]{fontenc}

\DeclareSymbolFont{cyrillic}{X2}{cmr}{m}{n}
\SetSymbolFont{cyrillic}{bold}{X2}{cmr}{bx}{n}
\DeclareMathSymbol{\CyLje}{\mathord}{cyrillic}{138}
\DeclareMathSymbol{\CyIe}{\mathord}{cyrillic}{170}
\DeclareMathSymbol{\CyBe}{\mathord}{cyrillic}{193}
\DeclareMathSymbol{\CyDe}{\mathord}{cyrillic}{196}
\DeclareMathSymbol{\CyZhe}{\mathord}{cyrillic}{198}
\DeclareMathSymbol{\CyVarZhe}{\mathord}{cyrillic}{199}
\DeclareMathSymbol{\CyI}{\mathord}{cyrillic}{200}
\DeclareMathSymbol{\CyEl}{\mathord}{cyrillic}{203}
\DeclareMathSymbol{\CyTse}{\mathord}{cyrillic}{214}
\DeclareMathSymbol{\CyChe}{\mathord}{cyrillic}{215}
\DeclareMathSymbol{\CySha}{\mathord}{cyrillic}{216}
\DeclareMathSymbol{\CyShcha}{\mathord}{cyrillic}{217}
\DeclareMathSymbol{\CyYer}{\mathord}{cyrillic}{218}
\DeclareMathSymbol{\CyFrontYer}{\mathord}{cyrillic}{220}
\DeclareMathSymbol{\CyE}{\mathord}{cyrillic}{221}
\DeclareMathSymbol{\CyYu}{\mathord}{cyrillic}{222}

\usepackage{lmodern}
\rmfamily
\DeclareFontShape{T1}{lmr}{b}{sc}{<->ssub*cmr/bx/sc}{}
\DeclareFontShape{T1}{lmr}{bx}{sc}{<->ssub*cmr/bx/sc}{}
\usepackage[cal=cm, scr=boondox, frak=euler]{mathalfa} % mathcal

\usepackage [english]{babel}
\usepackage{natbib}
\usepackage[ruled]{algorithm2e}
\usepackage{titlesec}

\renewcommand{\qedsymbol}{\null\nobreak\hfill\ensuremath{\square}}

\setcitestyle{round}

\hypersetup{
	colorlinks = true,
	citecolor = blue,
	linkcolor = blue  
}

\usepackage[nameinlink, noabbrev]{cleveref}

\AtBeginEnvironment{appendices}{\crefalias{section}{appendix}}

\creflabelformat{equation}{#2#1#3}
\crefrangelabelformat{equation}{#3#1#4--#5#2#6}


\titlelabel{\thetitle.\quad}

\newcolumntype{Y}{>{\arraybackslash}X}
\newcolumntype{Z}{>{\centering \arraybackslash}X}

\newcommand{\vect}[1]{\boldsymbol{\mathbf{#1}}}
\newcommand{\expect}{\mathbb{E}}

\newcommand{\defeq}{\vcentcolon=}
\DeclareMathOperator*{\argmax}{arg\,max}

\DeclareMathOperator{\Tr}{Tr}
\def\indicator{\mathbb{I}}

\newcommand{\appropto}{\mathrel{\vcenter{
			\offinterlineskip\halign{\hfil$##$\cr
				\propto\cr\noalign{\kern2pt}\sim\cr\noalign{\kern-2pt}}}}}

\newcommand{\widesim}[2][1.5]{
	\mathrel{\overset{#2}{\scalebox{#1}[1]{$\sim$}}}
}

\theoremstyle{plain}
\newtheorem{proposition}{Proposition}

\newtheorem{lemma}{Lemma}

\newtheoremstyle{bfremark}%
{}{}%
{}{}%
{\bfseries}{.}%
{ }%
{\thmname{#1}\thmnumber{ #2}\thmnote{ \normalfont (#3)}}
\theoremstyle{bfremark}
\newtheorem{assumption}{Assumption}
\newtheorem{definition}{Definition}

\crefname{lemma}{lemma}{lemmas}
\creflabelformat{lemma}{#2#1#3}
\crefrangelabelformat{lemma}{#3#1#4--#5#2#6}
\crefname{assumption}{assumption}{assumptions}
\creflabelformat{assumption}{#2#1#3}
\crefrangelabelformat{assumption}{#3#1#4--#5#2#6}
\crefrangelabelformat{definition}{#3#1#4--#5#2#6}

\newtheorem{example}{Example}

\crefname{example}{example}{examples}
\creflabelformat{example}{#2#1#3}
\crefrangelabelformat{example}{#3#1#4--#5#2#6}

\newtheorem*{remark}{Remark}

\newenvironment{proofenv}[1][\proofname]{
	\noindent
	\ifx&#1&%
		\textsc{proof.}
	\else
		\textsc{\MakeLowercase{#1}.}
	\fi
}{\qedsymbol \endproof}

\title{\fontsize{18}{18} \textsc{\textbf{multidimensional dynamic factor models}}}
\date{}

\begin{document}
%\onehalfspacing

% Title and abstract

\begin{abstract}
	This paper generalises dynamic factor models for multidimensional dependent data. In doing so, it develops an interpretable technique to study complex information sources ranging from repeated surveys with a varying number of respondents to panels of satellite images. We specialise our results to model microeconomic data on US households jointly with macroeconomic aggregates. This results in a powerful tool able to generate localised predictions, counterfactuals and impulse response functions for individual households, accounting for traditional time-series complexities depicted in the state-space literature. The model is also compatible with the growing focus of policymakers for real-time economic analysis as it is able to process observations online, while handling missing values and asynchronous data releases.
\end{abstract}

\begin{keyword}
	Dynamic Factor Models, Heterogeneous Agent Models, Multidimensional Time Series, Real-time Economic Analysis, State-space Models
\end{keyword}

\begin{frontmatter}
	\author[add1]{Matteo Barigozzi}
	\author[add2,add3]{Filippo Pellegrino \corref{cor1}}
	\address[add1]{University of Bologna}
	\address[add2]{Imperial College London}
	\address[add3]{London School of Economics and Political Science}
	
	\cortext[cor1]{Corresponding author: \texttt{f.pellegrino22@imperial.ac.uk}.}
\end{frontmatter}

\thispagestyle{empty} 

%%%%%%%%%%%%%%%%%%%%%%%%%%%%%%%%%%%%%%%%%%%%%%%%%%%%%%%%%%%%%%%%%%%%%%%%%%%%%%%%%%%%%%%%%%%%%%

% Main
\section{Introduction} \label{sec:introduction}

Nowadays, it is easy to find datasets with millions of observations and measurements taken over a broad range of time periods. However, complexity increases with the number of dimensions considered per period and, thus, not all datasets are created equal.

Tabular datasets are often easier to model than more abstract cases, including time series of satellite images and texts. This reduced complexity inspired the development of interpretable models with straightforward policy applications. For instance, tabular multivariate time series are commonly studied via impulse response functions and conditional forecasts to determine appropriate fiscal and monetary policy actions. Unstructured datasets have been mostly studied through less explainable models and, as a result, they are not as used for policy. This is a pity, given that they could be handy for a broad range of applications, including studying poverty through satellite images and predicting volatility from market-risk reports, as surveyed in \cite{mullainathan2017machine}. Agreeing with similar considerations, we propose a framework compatible with multidimensional dependent data that retains the explainability of traditional statistical models. 

Our approach suggests to reshape multidimensional data into a tabular multivariate time series with a peculiar vectorisation that accounts for temporal variations in the sample size and composition. Once the transformation is completed, we propose to model the resulting data on the basis of state-space methods \citep[e.g.,][]{harvey1990forecasting} and reduce the magnitude of the problem by extracting unobserved common components across multiple dimensions and time. In doing so, we manage to obtain an explainable technique flexible enough for handling complex datasets and capable of being linked with domain-specific concepts through identification schemes such as those in \cite{bai2015identification}. This dimensionality reduction technique can be interpreted as a generalisation of the one employed by dynamic factor models. As such, the origins of our methodology are rooted in psychometrics \citep{lawley1962factor} and time-series econometrics \citep[]{geweke1977DFM, forni2000generalized, forni2005generalized, forni2009opening, forni2001generalized, bernanke2005measuring, doz2012quasi, barigozzi2020quasi}. In light of that, we call it multidimensional dynamic factor model.

We specialise our manuscript for analysing microeconomic data on households and macroeconomic time series jointly. This problem is indeed multidimensional since, at each point in time $t$, we observe a survey containing $N_{t}$ households with $K$ characteristics of interest. Our modelling choice is compatible with economic theory and flexible enough to describe the characteristics of different groups of households, thus helping measuring income inequality.

Our use of repeated microeconomic surveys is different from traditional approaches: we neither pretreat the time-series cross-sectional data by transforming it into aggregate indices, nor model it via cross-sectional regressions with a linear trend predictor. Instead, it shares similarities with the approach in \cite{liu2021full}. We both use macroeconomic aggregates, households data and state-space modelling. However, we model everything in one step and within a single state space, while they use a two-step method in which the latent components are extracted from macroeconomic aggregates only. This allows us to have a system able to handle microeconomic complexities such as the temporal dynamics of each household. As a result, we make a better use of the data and model serial correlation in household income across groups of demographics. Besides, using cyclical and non-stationary latent components we distinguish between transitory and persistent determinants of household income. We are not aware of other papers handling similar complexities at once and refer to the introduction of \cite{liu2021full} for an in-depth survey of correlated articles.

Our empirical analysis is based on a large dataset containing macroeconomic aggregates from the Archival Federal Reserve Economic Data (ALFRED) and households information collected in the Consumer Expenditure (CE) Public Use Microdata (PUMD). Our empirical results highlight differences in household income among distinct demographics. In particular, we find that our MDFM is able to capture persistent parts of income linked with education and ethnicity. Besides, we show that our model is able to track the demographics surveyed in the CE PUMD before its official publication date, thus extending the findings in \cite{giannone2008nowcasting} and the scope of nowcasting to microeconomic problems.

\section{Methodology} \label{sec:methodology}

\subsection{Data processing} \label{sec:methodology:data}

This subsection illustrates our approach to process multidimensional multivariate dependent data.

\begin{assumption}[Data] \label{assumption:data}
	Let $\vect{H}_t \in \mathbb{R}^{N_t \times K}$ be a data matrix with $N_{t} > 0$ and $K>0$, and denote with $\vect{h}_{t}$ the $N_t K \times 1$ vectorisation of $\vect{H}_{t}$, for every point in time $t$. Besides, assume that $\vect{H}_t$ is a finite realisation of some stochastic process observed at any point in time $t \in \mathscr{T} \subseteq \{1, \ldots, T\}$ where $T \geq 1$.
\end{assumption}

\begin{remark}
	In our notation, $N_{t}$ denotes the number of subjects at each point in time. It is important to stress that we talk about ``subjects'' figuratively. Indeed, our definition is not restricted to individuals, but extends to any abstract thing with a data structure compatible with \cref{assumption:data}.
\end{remark}

\begin{example}[Time-series cross sections] In the case of time-series cross-sectional data
	\begin{align*}
		\vect{H}_t = \begin{pmatrix}
			H_{1,1,t} & \ldots & H_{1,K,t} \\
			\vdots & \ddots & \vdots \\
			H_{N_t,1,t} & \ldots & H_{N_t,K,t}
		\end{pmatrix}
	\end{align*}
	
	\noindent represents a cross section referring to time $t$ and
	\begin{align*}
		\vect{h}_t = \begin{pmatrix} H_{1, 1, t} & \ldots & H_{1, K, t} & \ldots & H_{N_t, 1, t} & \ldots & H_{N_t, K, t} \end{pmatrix}',
	\end{align*}
	
	\noindent where $N_t > 0$ is the number of cross-sectional observations for time $t$ and $K$ is the total number of covariates.  In social sciences, similar datasets generally represent complex surveys with a varying number of respondents. However, $\vect{H}_{t}$ could also represent more exotic data. For instance, the RGB representation of a satellite image taken at time $t$ with $N_{t}$ pixels.
\end{example}

\begin{example}[Time series]
	The banal case in which $\vect{H}_{t} \in \mathbb{R}^{N_{t}}$ gives a time series dataset. Indeed, $\vect{H}_t = (H_{1,t} \;\, \ldots, H_{N_{t},t})'$ for any $t \in \mathscr{T}$. The value taken by $N_{t}$ over all $t \in \mathscr{T}$ controls whether this dataset represents a univariate or a multivariate time series, and if it is fully observed.
\end{example}

With empirical problems involving this data structure, new (old) subjects can be added (removed) over time. This implies that the same subject can be observed in our dataset at different positions across multiple points in time. We account for this complexity by associating a subject-characteristic identifier to each entry of $\vect{h}_t$, for every point in time $t \in \mathscr{T}$ in which at least one characteristic is observed.

\begin{definition}[Identifiers] \label{def:identifiers}
	
	In order to allow for this complexity, we let
	\begin{align*}
		\mathscr{S}_t \defeq \{ f(i,t) : 1 \leq i \leq N_t K\},
	\end{align*}
	
	\noindent for any $t \in \mathscr{T}$. The function $f:\mathbb{N} \times \mathbb{N} \rightarrow \mathbb{N}$ is a convenient way for categorising different subjects. Indeed, we structure it to be equal to one when evaluated at $(1,1)$ and to have an incremental value for any pair referring to a new feature of the same subject, or to a new subject. This implies that $f(i_1, t_1)=f(i_2, t_2)$ if and only if $(i_1, t_1)$ and $(i_2, t_2)$ refer to the same subject-characteristic pair. Hence, we let
	\begin{align*}
		\mathscr{S} \defeq \bigcup_{t \in \mathscr{T}} \mathscr{S}_t
	\end{align*}
	
	\noindent be the set of all (observed) subject-characteristic pairs. For simplicity, we let $N \defeq \frac{|\mathscr{S}|}{K}$ be the number of unique subjects.
\end{definition}

\begin{remark}
	 Under \cref{assumption:data}, a minimum of one subject is observed across the whole sample and thus $N > 0$. Besides, note that $N$ is a natural number by construction.
\end{remark}

\noindent Finally, we build on \cref{def:identifiers} and reshape the data into a multivariate time series.

\begin{definition} \label{def:data_time_series}
	Indeed, we let $\vect{Y}_{t} \defeq \vect{W}_{t} \, \vect{h}_{t}$ to be a $NK \times 1$ vector of time series where $\vect{W}_{t}$ is a $NK \times N_{t} K$ matrix such that
	\begin{align*}
		W_{i, j, t} = 
		\begin{cases}
			1, & \text{if } f(j,t)=i \text{ and } i \in \mathscr{S}_{t}, \\
			0, &\text{otherwise},
		\end{cases}
	\end{align*}

	\noindent for any $t \in \mathscr{T}$, $1 \leq i \leq NK$ and $1 \leq j \leq N_{t} K$. In order to have a record of the non-missing entries of each $i$-th subject-characteristic pair, we also let $\mathscr{T}_{i} \subseteq \mathscr{T}$ to be the set of points in time in which it is observed. Interestingly, $\mathscr{T} = \bigcup_{i=1}^N \mathscr{T}_{i}$.
\end{definition}

\begin{remark}
	Note that by taking track of the observed datapoints via the $\mathscr{T}_{i}$ sets, we can distinguish between real zeros and missing values. Indeed, we do so in the estimation method described below. Moreover, due to \cref{assumption:data}, all $\mathscr{T}_{i}$ referring to a single subject are identical (i.e., when we observe a subject we measure all of its characteristics).
\end{remark}

\subsection{Multidimensional dynamic factor models} \label{sec:methodology:mdfm}

This subsection formalises our approach for modelling generic multidimensional multivariate data via dynamic factor models. This methodology is then specialised in \cref{sec:methodology:framework} focussing on economics.

A multidimensional dynamic factor model (MDFM) is a decomposition of the data class described in \cref{sec:methodology:data} into mutually orthogonal common and idiosyncratic components at all leads and lags.

\begin{assumption}[Generic MDFM]  \label{assumption:generic_model}
	Going forward, we assume that the model for any $\vect{Y}_{t}$ is the multidimensional dynamic factor model
	\begin{alignat*}{2}
		\vect{Y}_{t} &= \vect{B}(L) \vect{\Phi}_{t} + \vect{e}_{t}, \qquad &&\vect{e}_{t} \widesim{w.n.} N(\vect{0}_{NK \times 1}, \vect{R}), \\
		\vect{\Phi}_{t} &= \vect{C}(L) \vect{\Phi}_{t-1} + \vect{D} \vect{u}_{t}, \qquad &&\vect{u}_{t} \widesim{w.n.} N(\vect{0}_{r \times 1}, \vect{\Sigma}), 
	\end{alignat*}

	\noindent where $\vect{\Phi}_{t}$ denotes a vector of latent components (stationary and/or non-stationary) and for some positive definite covariance matrices $\vect{R}$ and $\vect{\Sigma}$.  The vector $\vect{\Phi}_{t}$ is $q$-dimensional with $1 \leq q \ll NK$, linked to the measurements via the matrix $\vect{B}(L)$ and with dynamics determined by $\vect{C}(L)$. Besides, $\vect{D}$ is $q \times r$ with $1 \leq r \leq q$.
\end{assumption}

\begin{remark}
	In this article, the common components are not restricted to be stationary.
\end{remark}

This model is extremely general and requires a set of restrictions in the parameters to be uniquely identified. This problem is analogous to the one observed with generic dynamic factor models (which are a particular case of MDFM) and it can be handled with the approaches proposed in \cite{bai2013principal} and \cite{bai2015identification}. However, if the empirical problem at hand requires the extraction of multiple factors from datasets with subjects observed once or very few times over the full sample, it becomes hard to do. In those cases, it is often convenient to construct the dataset using both time series and multidimensional data. In doing so, a minimal number of subjects (i.e., the number of time series) is observed for most/all points in time and can be used for defining more solid identifying restrictions. \Cref{sec:methodology:framework} follows this approach for proposing a specialised model for economic data.

As for the case of standard dynamic factor models, the MDFM can be estimated with an EM \citep{dempster1977maximum, rubin1982algorithms, shumway1982approach, watson1983alternative, banbura2014maximum, barigozzi2020quasi}, ECM \citep{meng1993maximum, pellegrino2020selecting, pellegrino2021factoraugmented} or ECME algorithm \citep{liu1994ecme}, as well as with Bayesian methods \citep{sarkka2013bayesian}. These techniques allow to have missing observations in the measurements, which are often found in the class of multidimensional data described in this manuscript.

\subsection{A microfounded dynamic factor model} \label{sec:methodology:framework}

We specialise our approach to model microeconomic data jointly with macroeconomic aggregates. This subsection introduces our empirical research question and the economics-informed restrictions we employ for identifying the MDFM.

We propose to use a MDFM for understanding the effect of expansions and recessions on individual US households. In particular, we aim to do so studying the sensitivity of their real income per head to changes in the business cycle (BC), while taking into account the differences that exist across demographic groups (both temporary and persistent). This is an important question for politicians and central bankers. Indeed, an accurate answer would allow to systematically target fiscal and monetary policies for addressing the needs of specific demographic groups.

We collect data on US households from the Consumer Expenditure (CE) Public Use Microdata (PUMD). This is a vast dataset containing information on consumers and their household, including demographic characteristics, income and expenditure figures. The data is collected by the Census Bureau for the Bureau of Labor Statistics in the Interview Survey and Diary Survey. We focus on the first -- which is the one describing major and/or recurring items -- to gather information on quarterly income before tax and descriptive characteristics at the household level.\footnote{Note that the BLS refers to households as consumer units (CUs). We use them as synonyms.} 

In particular, we use the FMLI and ITBI files published from 1990 to 2020 for constructing a quarterly dataset containing demographic and nominal income data.\footnote{We have decided to start from the 1990 file since the ITBI data was not available from 1981 to 1989. Note that the 1990 file also includes data referring to 1989 (from October).} We exclude the subset of households that has not provided enough information to be categorised under one or more of the demographic characteristics in \cref{tab:micro_data}, those whose attributes changed over time and the consumer units that have not provided any information on their income at all.\footnote{We do not exclude households whose income changed over time or with an incomplete income record, as long as we have at least one observation.} Moreover, we focus on prime working age urban consumer units (i.e., 25 to 54 years). The resulting dataset comprises a total of approximately 87,000 households.

\begin{table}[!t]
	\footnotesize
	\begin{tabularx}{\textwidth}{@{}lZZZ@{}}
		\toprule 
		Description & Mnemonic & Categorical & File \\
		\midrule
		Census region 							  & REGION                              & Y & FMLI \\
		College educated household    & EDUC\_HH                         & Y & FMLI \\
		Family size									  & FAM\_SIZE    				     & N & FMLI \\
		Family type  								 & FAM\_TYPE   					   & Y & FMLI \\
		Prime working age    			       & PRIME\_AGE 					& Y & FMLI \\
		Real household income per head (before tax)   & INCOME & N & FMLI and ITBI \\
		Urban consumers  	  				  & BLS\_URBN   				    & Y & FMLI \\
		White household                         & WHITE\_HH  					  & Y & FMLI \\
		\bottomrule
	\end{tabularx}
	\caption{Consumer Expenditure (CE) Public Use Microdata (PUMD) selection. The data is extracted from the FMLI and ITBI files published from 1990 to 2020. We deflate the nominal income per head data in the ITBI using the PCE price index in \cref{tab:macro_data}. Further details on the data construction are reported in \cref{appendix:pumd}. \\
	\textbf{Source:} Census Bureau for the Bureau of Labor Statistics, Bureau of Economic Analysis.}
	\label{tab:micro_data}
\end{table}

\begin{definition}[Groups] \label{def:groups}
	Define $\mathscr{G}$ as the Cartesian product of the household attributes on education and ethnicity in \cref{tab:micro_data}: a set with cardinality four such that each member is a unique combination of characteristics that identifies a specific demographic group. For simplicity, we refer to these groups in the order: (0, 0), (0, 1), (1, 0), (1, 1) whereas zero and one refers to the values taken by the binary variables EDUC\_HH and WHITE\_HH. Finally, we also let $\vect{\omega} \equiv \vect{\omega}(\mathscr{G})$ be the vector of integers denoting the number of households per group observed across all periods. 
\end{definition}

In addition to the microeconomic data, we also use macroeconomic aggregates. We first transform the nominal income figures obtained from the CE PUMD into real terms deflating them with the headline PCE price index -- keeping them at household level. Next, we merge the resulting real income figures with the macro dataset in \cref{tab:macro_data}. In order to perform these operations correctly, we download the macroeconomic series from the Archival Federal Reserve Economic Data (ALFRED) database and use the vintage released right after the 2020 CE PUMD Interview Survey's publication date.

\begin{definition}[Empirical data] \label{def:empirical_data}
	We then arrange the data to match the structure in \cref{def:data_time_series} and let
	\begin{align*}
		\vect{Y}_{t} = \big(\vect{X}_{t}' \;\, \vect{Z}_{1,t}'  \;\, \ldots \;\, \vect{Z}_{4,t}' \big)',
	\end{align*}
	
	\noindent where $\vect{X}_{t}$ denotes the vector of macroeconomic aggregates and each $\vect{Z}_{i,t}$ represents the vector of real income per head for all households in group $1 \leq i \leq 4$.
\end{definition}

\begin{remark}
	Note that every $\vect{Z}_{i,t}$ is $\omega_i \times 1$ dimensional. Since the CE PUMD is structured to survey the same household for a maximum of 4 quarters, the $\vect{Z}_{i,t}$ vectors are sparse. Besides, the missing observations in $\vect{Y}_{t}$ are handled as in \cref{def:data_time_series}. Finally, recall that this application focusses on the four demographic groups indicated in \cref{def:groups}.
\end{remark}

\noindent For simplicity, the macroeconomic aggregates are used in the same order reported in \cref{tab:macro_data}. Any within-group ordering for the households is equivalent for our MDFM. We collect them in ascending order, on the basis of the official NEWID identifier available in Consumer Expenditure Public Use Microdata.\footnote{Note that the last digit of the NEWID refers to the interview number and the previous ones identify the consumer units. As a result, we have not considered the last digit of NEWID to identify the households and determine the within-group ordering.}

Having shaped the data in the form prescribed in \cref{sec:methodology:data} we are now ready to specialise the MDFM for this household problem. Similarly to recent work on semi-structural models including \cite{hasenzagl2022a, hasenzagl2022b} and the empirical application in \cite{pellegrino2021factoraugmented}, we identify the model via economics-informed restrictions in order to extract interpretable unobserved components.

\begin{table}[!t]
	\footnotesize
	\begin{tabularx}{\textwidth}{@{}lZZ@{}}
		\toprule 
		Description \hspace{12.5em} & Mnemonic & Source \\
		\midrule		
		Real gross domestic product & GDPC1 & BEA \\ 
		Real personal consumption expenditures & PCECC96 & BEA \\
		Real gross private domestic investment & GPDIC1 & BEA \\
		Total nonfarm employment & PAYEMS & BLS \\
		Employment-population ratio & EMRATIO & BLS \\
		Unemployment rate & UNRATE & BLS \\
		Spot crude oil price (WTI) & WTISPLC & FRBSL \\
		Headline PCE & PCEPI & BEA \\
		\bottomrule
	\end{tabularx}
	\caption{Macroeconomic aggregates. The dataset is quarterly and includes all observations available in the vintage released right after the 2020 CE PUMD Interview Survey's publication date, starting from October 1989 (to be aligned with the 1990 ITBI). All series are downloaded and used in levels, except for the prices which are transformed in quarterly year-on-year percentage changes.\\ \textbf{Source:} Archival Federal Reserve Economic Data (ALFRED) database.}
	\label{tab:macro_data}
\end{table}

\begin{assumption}\label{assumption:empirical_mdfm}
	Formally, we let
	\begin{align*}
		\left( \hspace{-0.25em} \begin{array}{c} X_{1,t} \\ X_{2,t} \\ \vdots \\ X_{8,t} \\ \vect{Z}_{1,t} \\ \vect{Z}_{2,t} \\ \vect{Z}_{3,t} \\ \vect{Z}_{4,t} \end{array} \right) &= \left( \hspace{-0.25em} \begin{array}{c} \tau_{1,t} \\ \tau_{2,t} \\ \vdots \\ \tau_{8,t} \\ \tau_{9,t} \, \vect{\iota}_{\omega_1} \\ \tau_{10,t} \, \vect{\iota}_{\omega_2} \\ (\tau_{9,t} + \tau_{11,t}) \, \vect{\iota}_{\omega_3} \\ (\tau_{10, t} + \tau_{11, t}) \, \vect{\iota}_{\omega_4} \end{array} \right) + \left( \hspace{-0.25em}
		\begin{array}{cccc}
			1 \\
			\sum_{i=1}^{p} \Lambda_{1,i} L^{i-1} \\
			\vdots \\
			\sum_{i=1}^{p} \Lambda_{7,i} L^{i-1} \\
			\sum_{i=1}^{p} \Lambda_{8,i}  \, \vect{\iota}_{\omega_1} L^{i-1} \\
			\sum_{i=1}^{p} \Lambda_{9,i}  \, \vect{\iota}_{\omega_2} L^{i-1} \\
			\sum_{i=1}^{p} \Lambda_{10,i}  \, \vect{\iota}_{\omega_3} L^{i-1} \\
			\sum_{i=1}^{p} \Lambda_{11,i}  \, \vect{\iota}_{\omega_4} L^{i-1} \\
		\end{array} \right)
		\psi_{t} + \left( \hspace{-0.25em} \begin{array}{c} \xi_{1,t} \\ \xi_{2,t} \\ \vdots \\ \xi_{8,t} \\ \xi_{9,t} \\ \xi_{10,t} \\ \xi_{11,t} \\ \xi_{12, t} \end{array} \right) + \vect{e}_{t}
	\end{align*}
	
	\noindent where $\psi_{t}$ is a causal AR($p$) cycle denoting the business cycle; the $\tau$ denote smooth trends of order two modelled as in \citet[][ch. 8]{kitagawa1996smoothness}; $\xi_{1,t}, \ldots, \xi_{8+|\mathscr{G}|, t}$ are causal AR(1) latent components representing idiosyncratic noise; $\vect{\iota}$ denotes a vector of ones with length indicated in the subscript. Hereinafter, the number of lags $p$ is assumed being equal to $4$ (quarters).
\end{assumption}

\begin{remark}[Trends]
	Recall that a generic smooth trend $\underline{\tau}$ modelled as in \citet[][ch. 8]{kitagawa1996smoothness} is of order $k$ if $(1-L)^k \, \underline{\tau}$ is a white noise. Besides, note that the income figures share common trends. In particular: $\vect{\tau}_{9}$ models the persistent part of income for not college educated, not white households; $\vect{\tau}_{10}$ models the persistent part of income for not college educated, white households; $\vect{\tau}_{11}$ models the persistent offset of college educated households.
\end{remark}

\begin{remark}[CE PUMD data]
	\Cref{assumption:empirical_mdfm} implies that each household is modelled as a function of its own group and the dedicated parameters. In other words, all members of the $i$-th group are modelled via the same set of coefficients and latent factors, for every $1 \leq i \leq 4$. While the generic structure proposed in \cref{assumption:generic_model} could allow for a more disaggregate model, we do not have enough observations in the CE PUMD to do it. That being said, the model in \cref{assumption:empirical_mdfm} has quite a few advantages compared to these granular theoretical alternatives. Most importantly, it is less subject to idiosyncratic noise and due to the dimensionality reduction into group factors it is easier to interpret.
\end{remark}

The dynamics for the latent factors and the estimation method proposed for this model are illustrated in \cref{appendix:ecm}. The estimation is based on penalised quasi maximum likelihood estimation (PQMLE) and built on an ECM algorithm similar to the one employed in \cite{pellegrino2021factoraugmented}.

\section{Empirical results}

\subsection{Pre COVID-19 output}

We start analysing the results focussing on the pre COVID-19 period (1989 to 2019) and using a model estimated with the same cutoff. 

\begin{figure}[!t]
	\centering
	\includegraphics[width=\textwidth]{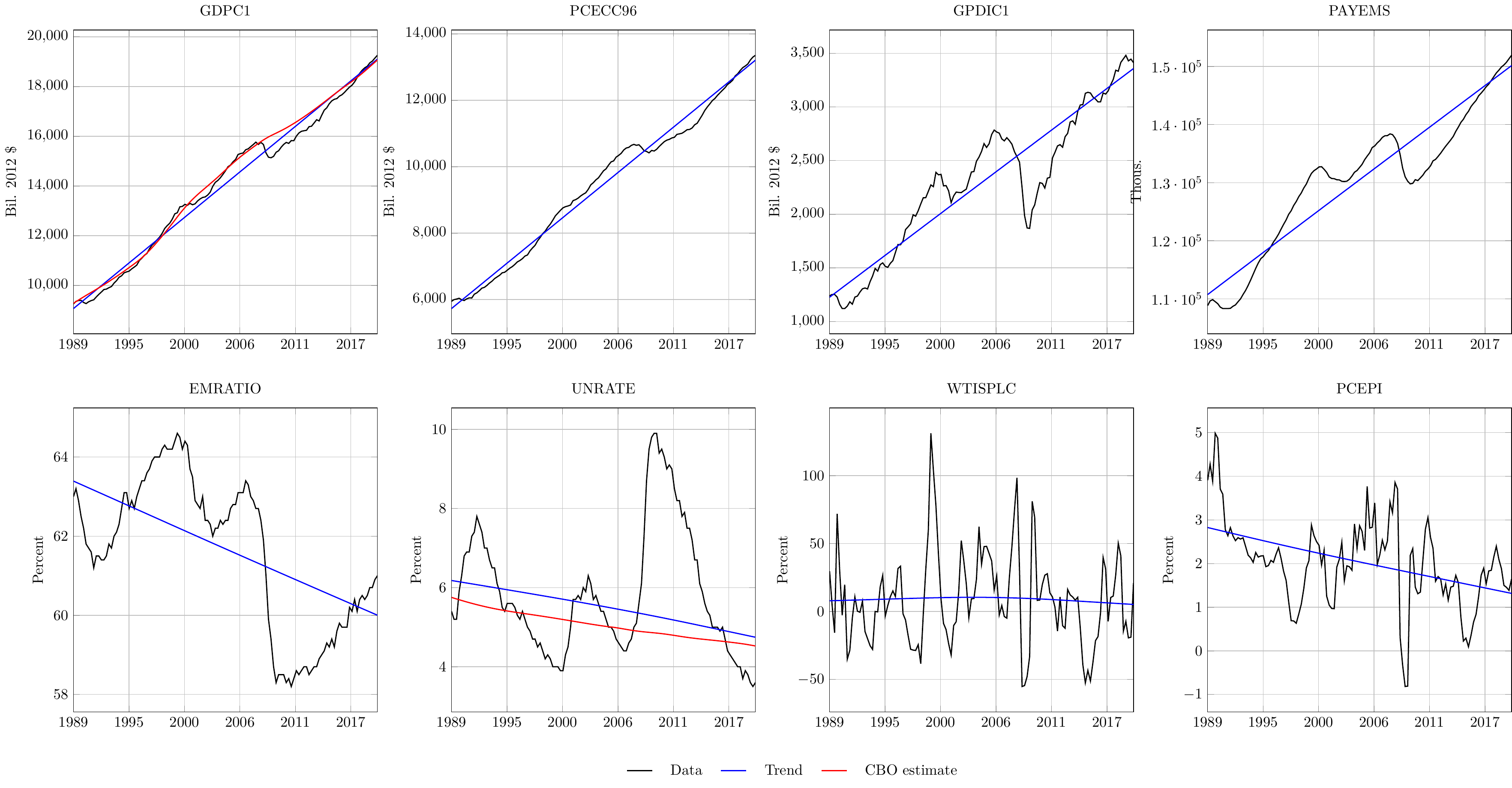}
	\caption{Macroeconomic data and trends. \\\textbf{Notes}: The model is estimated with quarterly data from October 1989 to December 2019. The congressional budget office estimates are aligned with the ALFRED vintage in \cref{tab:macro_data}.}
	\label{fig:macro_trends}
\end{figure}

\Cref{fig:macro_trends} reports the macroeconomic aggregates and their trends. The model uses them for describing the slow-moving and persistent component typical of economic time series. The difference between data and trend is the cycle. In the case of real GDP and unemployment rate, their trends are unobserved quantities of economic interest: the so-called potential output and non-accelerating inflation rate of unemployment (NAIRU). The Congressional Budget Office (CBO) publishes their own estimates for these objects which we use to benchmark ours. It is evident from \cref{fig:macro_trends} that there are strong differences only in the case of potential output. Indeed, our calculations imply a causal cycle with mean zero, whereas the CBO estimates a negative cycle for most periods. This is consistent with trend-cycle decompositions based purely on macroeconomic aggregates. Economic implications of this difference in view on potential output are discussed in \cite{hasenzagl2022a, hasenzagl2022b}.

\begin{figure}[!t]
	\centering
	\includegraphics[width=\textwidth]{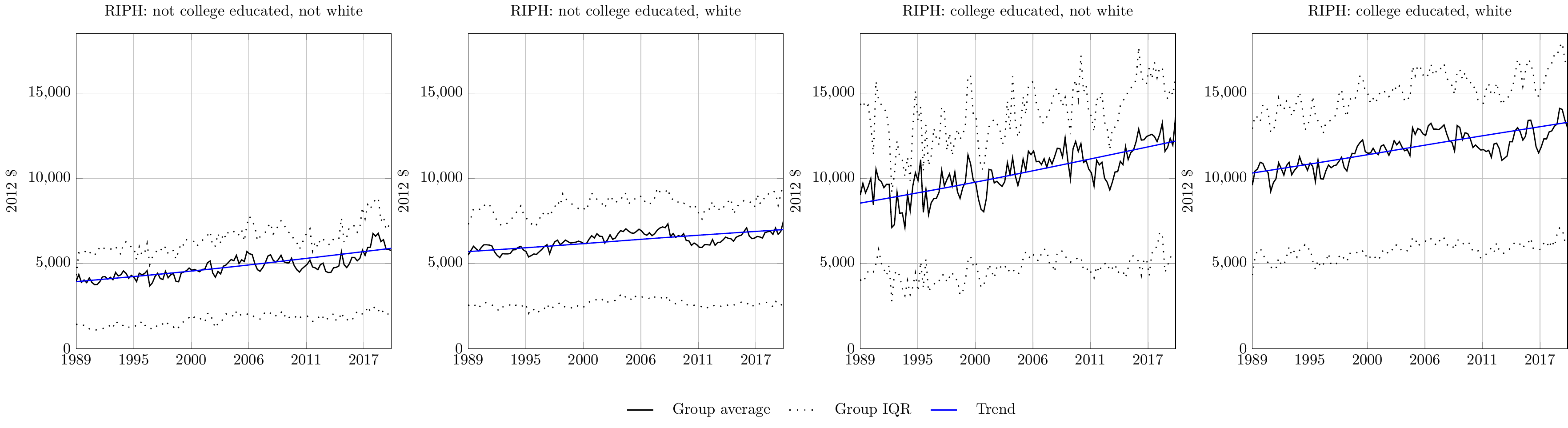}
	\caption{Microeconomic data and trends. \\\textbf{Notes}: The model is estimated with quarterly data from October 1989 to December 2019.}
	\label{fig:micro_trends}
\end{figure}

\Cref{fig:micro_trends} shows similar results for the income figures extracted from the CE PUMD. The main difference is that each subplot represents a group of households, not a single aggregate indicator. The demographic information is presented graphically through the following summary statistics: average, 25\% and 75\% quantiles. The trends do not refer to any specific household, but rather on the whole group. From a distributional standpoint, \cref{fig:micro_trends} shows four important points: most households have below-average income; a few individuals have disproportionate high revenues compared to the rest of their own demographic; white households are usually higher earners; college education increases the average income level. Our trend structure, further remarked after \cref{assumption:empirical_mdfm}, is flexible enough to accommodate for these features. Indeed, \cref{fig:micro_trends} show that white and college educated households have persistently higher trends.

\begin{figure}[!t]
	\centering
	\includegraphics[width=\textwidth]{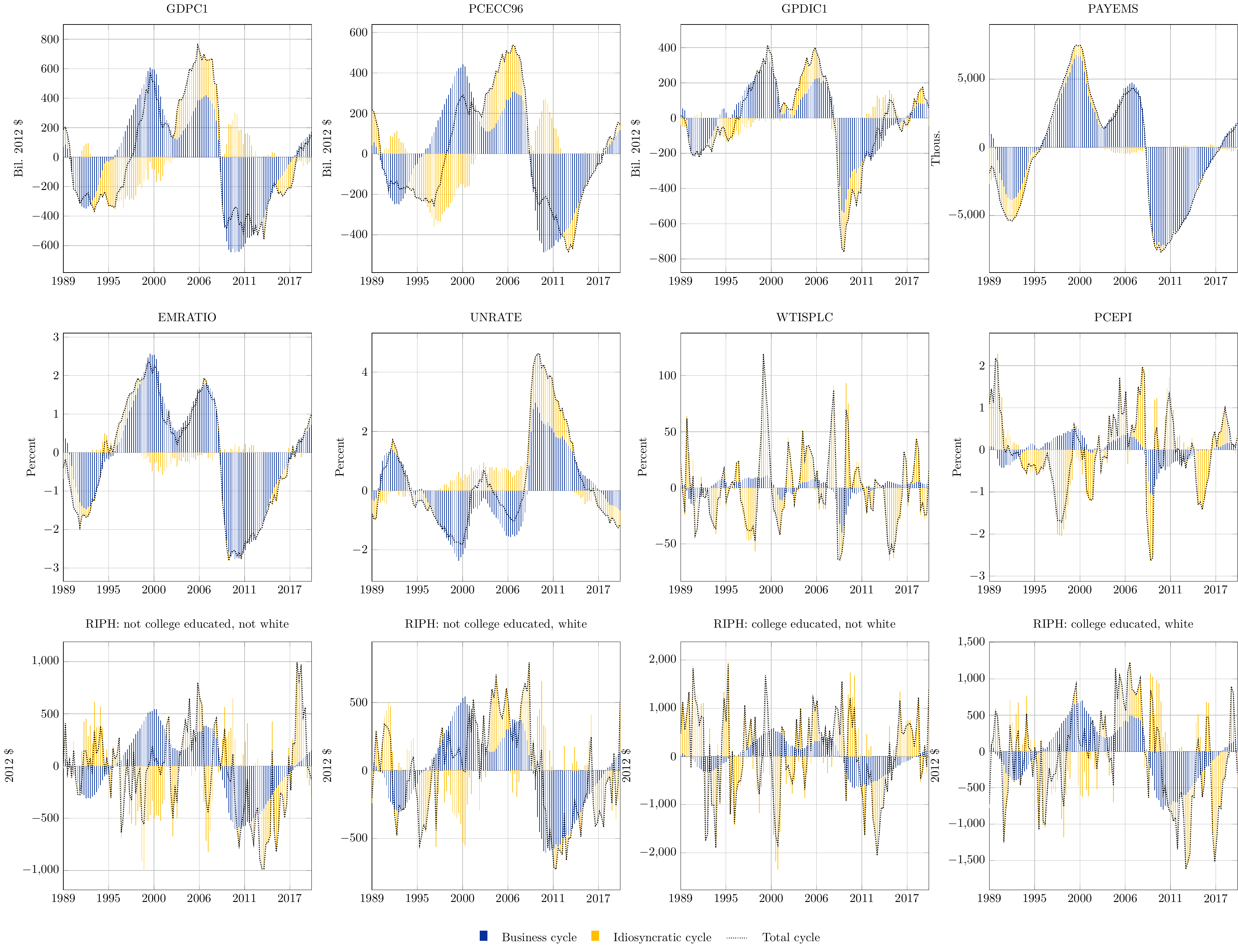}
	\caption{Historical decomposition of the cycles. \\\textbf{Notes}: The model is estimated with quarterly data from October 1989 to December 2019.}
	\label{fig:decomposition}
\end{figure}

\begin{figure}[!h]
		\centering
		\includegraphics[width=\textwidth]{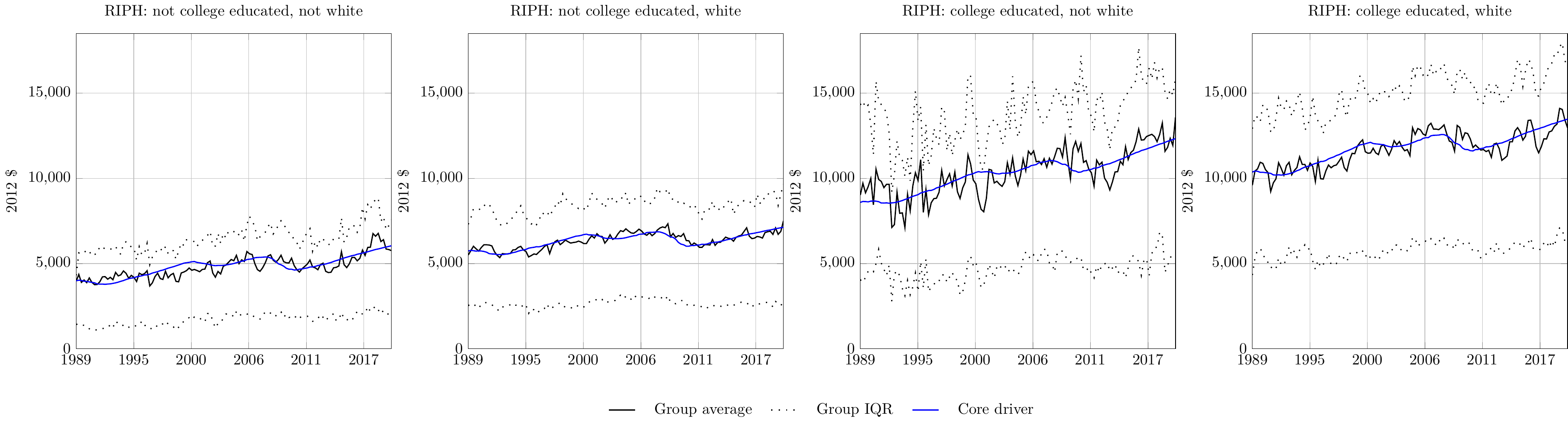}
		\caption{Core driver of RIPH computed as the sum of trend and business cycle. \\\textbf{Notes}: The model is estimated with quarterly data from October 1989 to December 2019.}
		\label{fig:core}
	\end{figure}

\Cref{fig:decomposition} breaks down the cycles to highlight commonalities and idiosyncrasies. The former are modelled through the business cycle and explain most of the cyclical fluctuations across macroeconomic aggregates and demographic groups. Idiosyncratic fluctuations, on the other hand, depict unique movements in specific macroeconomic indicators or groups. These are most prevalent for microeconomic data. Indeed, while the effect of the business cycle is comparable across demographics, each group exhibits distinct idiosyncratic patterns. \Cref{fig:core} builds on this further reporting the core drivers of the demographic groups: the sum between their trends and business cycles. Stripping out the idiosyncratic cycle helps visualising the crucial parts of real household income. Indeed, the resulting series is less impacted by outliers and noise.

\subsection{COVID-19 dataflow}

We now focus on the dataflow from January 2020 to March 2021 for studying the impact of COVID-19 on our estimates for the demographic groups. Throughout this subsection, we keep using the coefficients estimated with data from 1989 to 2019 to avoid altering the business cycle periodicity with a non traditional recession.

\begin{figure}[!t]
	\centering
	\includegraphics[width=\textwidth]{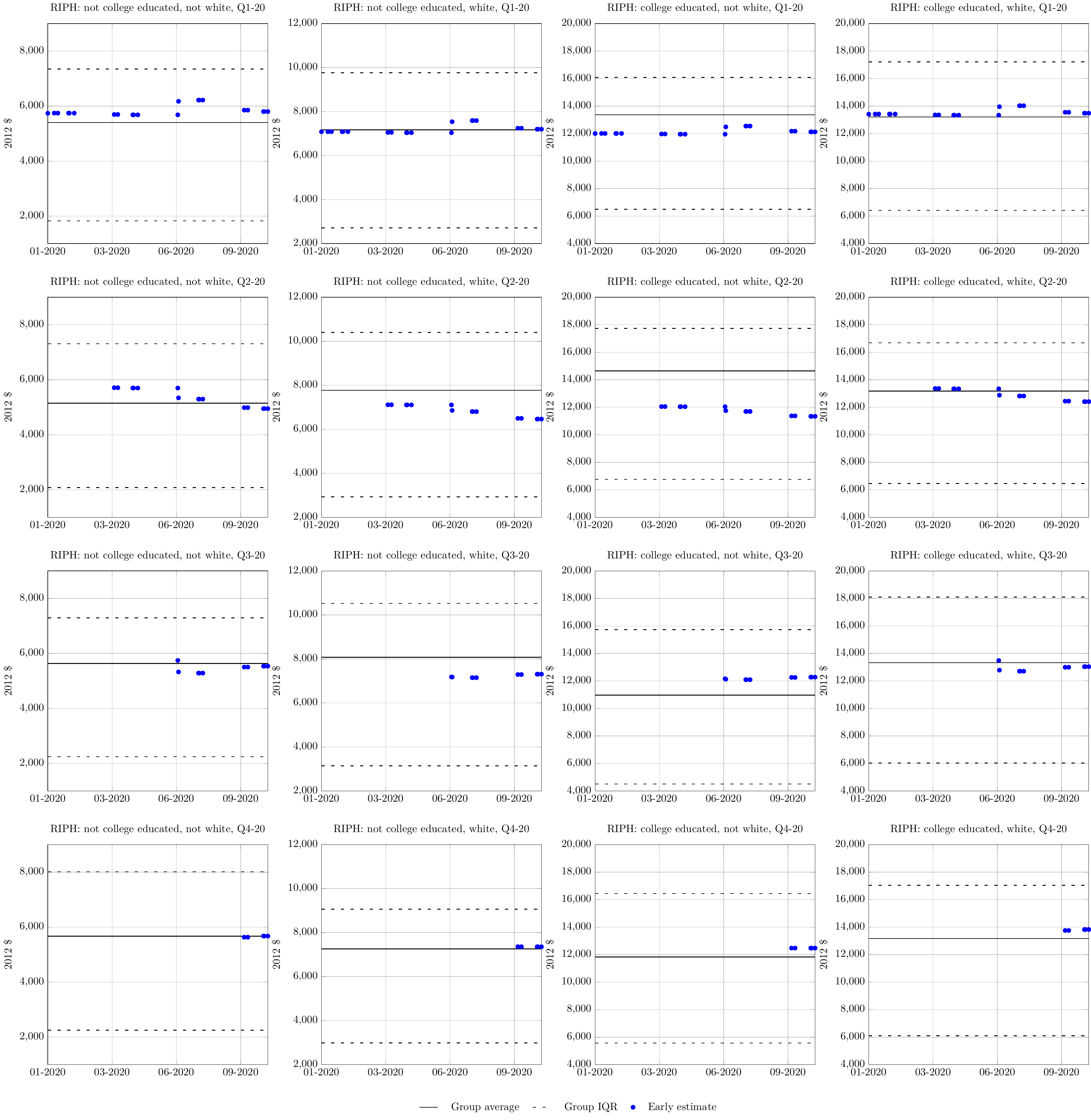}
	\caption{COVID-19 period dataflow. \\\textbf{Notes}: The model is estimated with quarterly data from October 1989 to December 2019. The dataflow contains macroeconomic releases from January 2020 to March 2021. The x-axis reports release dates.}
	\label{fig:nowcasting}
\end{figure}

Before getting into the results, it is important to mention that the CE PUMD files are released in one block for the whole year and with a large delay from their reference period. Indeed, it is usually possible to access data in the Interview Survey only after 3 months from the end of its reference year. For instance, the 2020 CE PUMD was released at the end of March 2021. However, extending our information set with more timely macroeconomic data we can compute early estimates.

We process the hereinbefore mentioned dataflow in a pseudo real-time fashion and generate early estimates for the microeconomic data at each release. In particular, we produce backcasts and nowcasts: forecasts referring to the previous and current reference quarters \citep{giannone2008nowcasting}. Given that the CE PUMD data is released one go, we keep backcasting previous quarters until the publication date. \Cref{fig:nowcasting} reports the results for each demographic of interest and, for simplicity, denotes backcasts and nowcasts as ``early estimates''. Overall, these predictions fluctuate closely to the ex-post group averages. This happens almost immediately and, thus, the expanding macroeconomic information set does not have a strong impact. The early estimates for the second quarter are the most distant from the ex-post group averages. This is not surprising since the strongest effect of COVID-19 on economic data was measured in that quarter. We can also see that sentiment (and expectations) became increasingly negative since after March, when the World Health Organization (WHO) declared COVID-19 a pandemic.

\section{Concluding remarks}

This article proposes to generalise Dynamic Factor Models to multidimensional data. The resulting framework is flexible enough to accommodate complex datasets ranging from surveys with varying number of respondents to time series of satellite images. However, it retains the interpretability typical of traditional factor models. 

We specialise our approach to model macroeconomic aggregates jointly with microeconomic data on household income. In this analysis, we study the effect of college education and ethnicity on the household income levels. In doing so, we find that our model is capable of recognising differences among demographics consistent with well-known stylised facts. Indeed, it finds that college education has a positive and persistent effect on household income and that white consumer units usually have higher earnings. We also explore the cyclical fluctuations in the data and highlight the heterogeneity among demographics.

Finally, realising that CE PUMD files are released with a large delay from their reference period, we show how to track them in real time focussing on the macroeconomic dataflow between January 2020 and March 2021. This is in line with the nowcasting literature \citep{giannone2008nowcasting} and, to the best of our knowledge, the first attempt to perform a similar exercise on microeconomic data.

% References
\bibliographystyle{abbrvnat}
\bibliography{biblio}

% Appendix
\clearpage
\begin{subappendices}
	
	\section{CE PUMD} \label{appendix:pumd}
	The demographic characteristics in \cref{tab:micro_data} are constructed at the household level. The following paragraphs give further details on each variable.
	
	\begin{itemize}
		\item \emph{Census region}: Census Bureau classification for US regions (1	Northeast, 2	Midwest, 3 South, 4	West). We use it for excluding CUs that moved across the US during the sampling period.
		\item \emph{College educated households}: describes the highest level of education of the reference person and spouse (if any). It is a dummy variable equal to 1 for CUs in which the highest education level is, at least, at an undergraduate level and 0 otherwise.
		\item \emph{Family size}: Number of family members. We use it for computing real household income per head (before tax).
		\item \emph{Family type}: Family categorisation. We use it for determine whether we there is a spouse to consider when constructing the other variables in this appendix.
		\item \emph{Prime working age}: dummy variable equal to 1 for CUs with average age between 25 and 55 years (excluded) and 0 otherwise. The average age is computed by taking the sample mean between the age of the reference person and spouse (if any). We use it for excluding non prime working age households.
		\item \emph{Urban consumers}: dummy variable equal to 1 for urban CUs and 0 otherwise. We use it for excluding rural CUs.
		\item \emph{White household}: dummy variable equal to 1 for white CUs and 0 otherwise.
	\end{itemize}

	The nominal income per head (before tax) is computed by constructing total nominal income from the ITBI files and dividing it for the number of CUs members in the FMLI files. The identifiers or Universal Classification Code (UCC) for each single income component used for computing this total are summarised in \cref{tab:UCCs}.
	
	\begin{table}[!t]
		\footnotesize
		\begin{tabularx}{\textwidth}{@{}lZ@{}}
			\toprule 
			Releases & Universal Classification Codes (UCCs) \\
			\midrule
			\multirow{3}{*}{1990 to 2003} & 900000, 900010, 900020, 900030, 900040, 900080, 900050, 900060, \\
																 & 900070, 900100, 900110, 900090, 900120, 900150, 900131, 900132, \\
																 & 800700I, 800710I, 900140 \\
			\midrule			 
			\multirow{3}{*}{2004 to 2012} & 900000, 900010, 900020, 900030, 900040, 900080, 900050, 900060, \\
																 & 900070, 900100, 900110, 900090, 900120, 900150, 900131, 900132, \\
																 & 800700, 800710, 900140 \\
			\midrule
			\multirow{2}{*}{2013 to 2020} & 900000, 900160, 900030, 900170, 900180, 900190, 900200, \\
															     & 900090, 900120, 900150, 900210, 800700, 800710, 900140 \\
			\bottomrule
		\end{tabularx}
		\caption{Universal Classification Codes (UCCs) used for computing nominal income. \\
			\textbf{Source:} Census Bureau for the Bureau of Labor Statistics.}
		\label{tab:UCCs}
	\end{table}

	\clearpage
	\section{ECM algorithm} \label{appendix:ecm}
	This appendix develops an ECM algorithm \citep{meng1993maximum} to estimate the MDFM in \cref{sec:methodology:framework}. The design builds on \citet[Appendix C]{pellegrino2020selecting} and \citet[Appendix A]{pellegrino2021factoraugmented}. This manuscript uses the ``hat'' symbol to denote the estimated coefficients, an $s$ subscript to indicate the sample size and a $k$ superscript for the ECM iteration.
	
	\subsection{State-space representation} \label{appendix:ecm:ssm}
	
	Recall that 
	\begin{alignat*}{2}
		\vect{Y}_{t} &= \vect{B}(L) \vect{\Phi}_{t} + \vect{e}_{t}, \qquad &&\vect{e}_{t} 	\widesim{w.n.} N(\vect{0}_{NK \times 1}, \vect{R}), \\
		\vect{\Phi}_{t} &= \vect{C}(L) \vect{\Phi}_{t-1} + \vect{D} \vect{u}_{t}, \qquad &&\vect{u}_{t} \widesim{w.n.} N(\vect{0}_{r \times 1}, \vect{\Sigma}).
	\end{alignat*}

	\noindent The matrices $\vect{C}(L)$ and $\vect{D}$ are sparse and their non-zero entries are such that
	\begin{align*}
		\vect{C}(L) = &\left( \hspace{-0.25em} 
		\begin{array}{c | ccc | ccccc | c}
			2 \vect{I}_{7+|\mathscr{G}|} & \cdot & \cdot  & \cdot & \cdot & \cdot & \cdot & \cdot & \cdot & - \vect{I}_{7+|\mathscr{G}|} \\
			\hline
			\cdot & \pi_{1} & \cdot & \cdot & \cdot & \cdot & \cdot & \cdot & \cdot & \cdot \\
			\cdot & \cdot & \ddots & \cdot & \cdot & \cdot & \cdot & \cdot & \cdot & \cdot \\
			\cdot & \cdot & \cdot & \pi_{8+|\mathscr{G}|} & \cdot & \cdot & \cdot & \cdot & \cdot & \cdot \\
			\hline
			\cdot & \cdot & \cdot & \cdot & \pi_{8+|\mathscr{G}|+1} & \pi_{8+|\mathscr{G}|+2} & \ldots & \pi_{8+|\mathscr{G}|+p-1} & \pi_{8+|\mathscr{G}|+p} & \cdot \\
			\cdot & \cdot & \cdot & \cdot & 1 & \cdot 	& \cdot & \cdot & \cdot & \cdot \\
			\cdot & \cdot & \cdot & \cdot & \cdot & 1 	& \cdot & \cdot & \cdot & \cdot \\
			\cdot & \cdot & \cdot & \cdot & \cdot & \cdot & \ddots & \cdot & \cdot & \cdot \\
			\cdot & \cdot & \cdot & \cdot & \cdot & \cdot & \cdot & 1 & \cdot & \cdot \\
			\hline
			\vect{I}_{7+|\mathscr{G}|} & \cdot & \cdot  & \cdot & \cdot & \cdot & \cdot & \cdot & \cdot & \cdot \\
		\end{array} \right) \\[-1em]
		&\hspace{0.85em} \begin{array}{c ccc ccccc c}
			\multicolumn{1}{c}{\hspace{-.1em}\underbrace{\hspace{2.8em}}_{q \times 7+|\mathscr{G}|}} & 
			\multicolumn{3}{c}{\hspace{-.2em}\underbrace{\hspace{6.5em}}_{q \times 8+|\mathscr{G}|}} & \multicolumn{5}{c}{\hspace{-.3em}\underbrace{\hspace{18.95em}}_{q \times p}} & \multicolumn{1}{c}{\hspace{-.3em}\underbrace{\hspace{3em}}_{q \times 7+|\mathscr{G}|}}
		\end{array}
	\end{align*}

	\noindent and
	\begin{align*}
		\vect{D} = &\left( \hspace{-0.25em} 
		\begin{array}{c | c | c }
			\vect{I}_{7+|\mathscr{G}|} & \cdot & \cdot \\
			\hline
			\cdot & \vect{I}_{8+|\mathscr{G}|} & \cdot \\
			\hline
			\cdot & \cdot & \phantom{1} 1 \phantom{1} \\
			\hline
			\cdot & \cdot & \cdot
		\end{array} \right) \\[-1em]
		&\hspace{0.85em} \begin{array}{c c c}
		\multicolumn{1}{c}{\hspace{-.3em}\underbrace{\hspace{2.4em}}_{q \times 7+|\mathscr{G}|}} & 
		\multicolumn{1}{c}{\hspace{-.5em}\underbrace{\hspace{2.4em}}_{q \times 8+|\mathscr{G}|}} & \multicolumn{1}{c}{\hspace{-.3em}\underbrace{\hspace{1.8em}}_{q \times 1}}
	\end{array}
	\end{align*}

	\noindent where $\vect{\pi}$ is a $8+|\mathscr{G}|+p \times 1$ vector of finite real parameters which ensures that the cyclical components are causal. Due to the structure of $\vect{C}$ and $\vect{D}$, it follows that $r = 16+2|\mathscr{G}|$ and $q=22+3|\mathscr{G}|+p$. The measurement coefficient matrix $\vect{B}(L)$ is also sparse and its non-zero entries are
	\begin{footnotesize}
		\begin{align*}
			&\left( \hspace{-0.25em} 
			\begin{array}{ccccccc | ccccccc | cccc | c}
				1  		 & \cdot & \cdot & \cdot & \cdot & \cdot & \phantom{ } \cdot \phantom{ } & 1 		 & \cdot & \cdot & \cdot & \cdot & \cdot & \cdot  & 1 & \cdot & \ldots & \cdot & \cdot  \\
				\cdot & 1	     & \cdot & \cdot & \cdot & \cdot & \cdot & \cdot & 1        & \cdot & \cdot & \cdot & \cdot & \cdot & \Lambda_{1,1} & \Lambda_{1,2} & \ldots & \Lambda_{1,p} & \cdot \\
				\cdot & \cdot & \ddots 		  & \cdot & \cdot & \cdot & \cdot & \cdot & \cdot & \ddots       & \cdot & \cdot & \cdot & \cdot & \vdots & \vdots & \vdots & \vdots & \cdot  \\
				\cdot & \cdot & \cdot  & 1		  & \cdot & \cdot & \cdot & \cdot & \cdot & \cdot & 1		 & \cdot & \cdot & \cdot & \Lambda_{7,1} & \Lambda_{7,2} & \ldots & \Lambda_{7,p} & \cdot  \\
				\cdot & \cdot & \cdot  & \cdot & 1 		  & \cdot & \cdot & \cdot & \cdot & \cdot & \cdot & 1		  & \cdot & \cdot & \Lambda_{8,1} \, \vect{\iota}_{\omega_{1}}  & \Lambda_{8,2} \, \vect{\iota}_{\omega_{1}} & \ldots & \Lambda_{8,p} \, \vect{\iota}_{\omega_{1}} & \cdot  \\
				\cdot & \cdot & \cdot  & \cdot & \cdot & \ddots  	  & \cdot & \cdot & \cdot & \cdot & \cdot & \cdot  & \ddots & \cdot & \vdots & \vdots & \vdots & \vdots & \phantom{1} \cdot \phantom{1} \\
				\cdot & \cdot & \cdot  & \cdot & \cdot & \cdot & 1        & \cdot & \cdot & \cdot & \cdot & \cdot & \cdot & 1         & \Lambda_{7+|\mathscr{G}|,1} \, \vect{\iota}_{\omega_{|\mathscr{G}|}}& \Lambda_{7+|\mathscr{G}|,2} \, \vect{\iota}_{\omega_{|\mathscr{G}|}} & \ldots & \Lambda_{7+|\mathscr{G}|,p} \, \vect{\iota}_{\omega_{|\mathscr{G}|}} & \cdot 
			\end{array} \right).\\[-1em]
			&\hspace{0.8em} \begin{array}{ccccccc ccccccc cccc c}
				\multicolumn{7}{c}{\underbrace{\hspace{11.35em}}_{NK \times 7+|\mathscr{G}|}} & 
				\multicolumn{7}{c}{\hspace{-0.5em}\underbrace{\hspace{11.4em}}_{NK \times 8+|\mathscr{G}|}} & \multicolumn{4}{c}{\hspace{-0.5em}\underbrace{\hspace{21em}}_{NK \times p}} & \multicolumn{1}{c}{\hspace{-1.45em}\underbrace{\hspace{2.2em}}_{NK \times 7+|\mathscr{G}|}}
			\end{array}
		\end{align*}
	\end{footnotesize}
	
	\noindent As a result,
	\begin{align*}
		&\vect{\Phi}_{t} \defeq \left( \begin{array}{ccc | ccc | cccc | ccc} \tau_{1,t} & \ldots & \tau_{7+|\mathscr{G}|,t} & \xi_{1,t} & \ldots & \xi_{8+|\mathscr{G}|, t} & \psi_{t} & \psi_{t-1} & \ldots & \psi_{t-p+1} & \tau_{1,t-1} & \ldots & \tau_{7+|\mathscr{G}|,t-1} \end{array} \right)'. \\[-1em]
		&\hspace{3.1em} \begin{array}{ccc ccc cccc ccc}
			\multicolumn{3}{c}{\underbrace{\hspace{7.1em}}_{7+|\mathscr{G}| \times 1}} &
			\multicolumn{3}{c}{\hspace{-0.26em}\underbrace{\hspace{7.2em}}_{8+|\mathscr{G}| \times 1}} &
			\multicolumn{4}{c}{\hspace{-0.3em}\underbrace{\hspace{9.5em}}_{p \times 1}} &	\multicolumn{3}{c}{\hspace{-0.32em}\underbrace{\hspace{9em}}_{7+|\mathscr{G}| \times 1}}
		\end{array}
	\end{align*}
	
	\begin{assumption}[Initial conditions] \label{assumption:ssm_init_cond}
		Given that we observe data at time $t=1$, we further assume that $\vect{\Phi}_{0} \widesim{w.n.} N(\vect{\mu}_{0}, \vect{\Omega}_{0})$ for some finite vector of real parameters $\vect{\mu}_{0}$ and a positive definite covariance matrix $\vect{\Omega}_{0}$. The latter is assumed to be sparse and such that the only entries allowed to differ from zero are those denoting the initial auto-covariances of each state.
	\end{assumption}
	
	\begin{remark}[Non-zero entries of $\vect{\Omega}_{0}$]
		In other words, the entries of $\vect{\Omega}_{0}$ that are allowed to differ from zero are those with coordinates $(i,j)$ in the union of the following sets:
		\begin{itemize}
			\item $\{(i,j) : i=j \text{ and } 1 \leq i < r\}$;
			\item $\{(i,j) :r \leq i < r+p \text{ and } r \leq j < r+p\}$.
		\end{itemize}
	\end{remark}
	
	\subsection{Estimation}
	
	This manuscript builds on the theoretical results in \cite{barigozzi2020quasi} and estimates the model via quasi penalised maximum likelihood estimation (PQMLE) by  considering $\vect{\Sigma}$ as a diagonal matrix and $\vect{R} = \varepsilon \, \vect{I}_{NK}$ for a small positive $\varepsilon$.\footnote{We set $\varepsilon = 10^{-2}$.} Formally, this implies that the free parameters to estimate are
	\begin{align*}
		&\vect{\vartheta} \defeq \left( \hspace{-0.1em} \begin{array}{cccccccc} \vect{\mu}_0' & \text{vech}(\vect{\Omega}_0)' & \text{vec}(\vect{\Lambda})' & \vect{\pi}' & \Sigma_{1,1} & \Sigma_{2,2} & \ldots & \Sigma_{r, r} \end{array} \right)'.
	\end{align*}
	
	The estimation is performed with an ECM algorithm: an optimisation method that repeats the operations in \cref{def:ecm_routine} until it reaches convergence. 
	
	\begin{definition}[ECM estimation routine] \label{def:ecm_routine}
		At any $k+1 > 1$ iteration, the ECM algorithm computes the vector of coefficients
		\begin{align*}
			\vect{\hat{\vartheta}}_{s}^{k+1}(\vect{\gamma}) \defeq \argmax_{\underline{\vect{\vartheta}} \, \in \, \mathscr{R}} \, \expect \left [\mathcal{L}(\underline{\vect{\vartheta}} \,|\, \vect{Y}_{1:s}, \vect{\Phi}_{1:s}) \,|\, \mathscr{Y}(s), \, \vect{\hat{\vartheta}}_{s}^{k}(\vect{\gamma}) \right] - \expect \left [\mathcal{P}(\underline{\vect{\vartheta}}, \vect{\gamma}) \,|\, \mathscr{Y}(s), \, \vect{\hat{\vartheta}}_{s}^{k}(\vect{\gamma}) \right],
		\end{align*}
		
		\noindent where $\mathscr{R}$ denotes the region in which the AR cycles (common and idiosyncratic) are causal, $\mathscr{Y}(s)$ is the information set available at time $s$,
		\begin{align} \label{eq:complete_loglik}
			\mathcal{L}(\underline{\vect{\vartheta}} \,|\, \vect{Y}_{1:s}, \vect{\Phi}_{1:s}) \simeq &-\frac{1}{2}\ln|\underline{\vect{\Omega}_0}| - \frac{1}{2}\Tr\Big[\underline{\vect{\Omega}_0}^{-1} (\vect{\Phi}_0 - \underline{\vect{\mu}_0})(\vect{\Phi}_0 - \underline{\vect{\mu}_0})'\Big] \\
			&-\frac{s}{2}\ln|\underline{\vect{\Sigma}}| - \frac{1}{2} \Tr \Big[\sum_{t=1}^s \underline{\vect{\Sigma}}^{-1} (\vect{\Phi}_{1:r, t} - \underline{\vect{C}}_{\,*} \vect{\Phi}_{t-1})(\vect{\Phi}_{1:r, t} - \underline{\vect{C}}_{\,*} \vect{\Phi}_{t-1})'\Big] \nonumber \\
			&-\frac{s}{2}\ln|\underline{\vect{R}}| - \frac{1}{2} \Tr\Big[\sum_{t=1}^s \underline{\vect{R}}^{-1} (\vect{Y}_t - \underline{\vect{B}}\vect{\Phi}_{t})(\vect{Y}_t - \underline{\vect{B}}\vect{\Phi}_{t})'\Big], \nonumber
		\end{align}
		
		\noindent $\underline{\vect{C}}_{\,*} \equiv \underline{\vect{C}}_{\,1:r, 1:q}$ and the underlined coefficients denote the parameters implied by $\underline{\vect{\vartheta}}$. The function in \cref{eq:complete_loglik} is the so-called complete-data (i.e., fully observed data and known latent states) log-likelihood. Besides,
		\begin{align*}
			\mathcal{P}(\underline{\vect{\vartheta}}, \vect{\gamma}) \defeq +\frac{1-\alpha}{2} &\left( \big\Vert \underline{\vect{\pi}}_{\,1:8+|\mathscr{G}|} \, \vect{\Gamma}(\vect{\gamma}, 1)^{\frac{1}{2}} \big\Vert_{\text{F}}^2 + \big\Vert \underline{\vect{\pi}}_{\,8+|\mathscr{G}|+1:8+|\mathscr{G}|+p}' \, \vect{\Gamma}(\vect{\gamma}, p)^{\frac{1}{2}} \big\Vert_{\text{F}}^2 + \big\Vert \underline{\vect{\Lambda}} \, \vect{\Gamma}(\vect{\gamma}, p)^{\frac{1}{2}} \big\Vert_{\text{F}}^2 \right) \\
			+ \frac{\alpha}{2} &\left( \big\Vert \underline{\vect{\pi}}_{\,1:8+|\mathscr{G}|} \, \vect{\Gamma}(\vect{\gamma}, 1) \big\Vert_{1,1} + \big\Vert \underline{\vect{\pi}}_{\,8+|\mathscr{G}|+1:8+|\mathscr{G}|+p}' \, \vect{\Gamma}(\vect{\gamma}, p) \big\Vert_{1,1} + \big\Vert \underline{\vect{\Lambda}} \, \vect{\Gamma}(\vect{\gamma}, p) \big\Vert_{1,1} \right)
		\end{align*}
		
		\noindent is a version of the elastic-net penalty in \cite{pellegrino2020selecting} in which, for any $l \in \mathbb{N}$,
		\begin{align*}
			\vect{\Gamma}(\vect{\gamma}, l) \defeq \rho \begin{pmatrix}
				1 & 0 &\ldots & 0 \\
				0 &\beta &\ldots & 0 \\
				\vdots &\ddots &\ddots &\vdots \\
				0 &\ldots &\ldots & \beta^{l-1} \end{pmatrix},
		\end{align*}
		
		\noindent $\rho \geq 0$, $0 \leq \alpha \leq 1$ and $\beta \geq 1$ are hyperparameters included in $\vect{\gamma}$. The state-space coefficients for the first iteration are initialised as in \cref{appendix:ecm:init}.
	\end{definition}
	
	\begin{assumption}[Convergence]
		The ECM algorithm is considered to be converged when the estimated coefficients do not significantly change in two subsequent iterations. This is done by computing the absolute relative change per parameters and comparing at the same time the median and $95^{th}$ quantile with a fixed tolerance of $10^{-3}$ and $10^{-2}$ respectively.
	\end{assumption}
	
	The operation in \cref{def:ecm_routine} is performed in two steps: the so-called E-step and CM-step. The E-step computes the expectations in \cref{def:ecm_routine}, while the CM-step conditionally maximises them with respect to the free parameters. 
	
	We write down the E-step on the basis of the output of a Kalman smoother compatible with incomplete data, as originally proposed in \cite{shumway1982approach} and \cite{watson1983alternative}. For that, we use the notation in \cref{def:kalman}.
	
	\begin{definition}[Kalman smoother output] \label{def:kalman}
		The hereinbefore mentioned Kalman smoother output is
		\begin{align*}
			&\vect{\hat{\Phi}}_{t} \defeq \expect \, \Big[\vect{\Phi}_t \,|\, \mathscr{Y}(s), \vect{\hat{\vartheta}}_{s}^{k}(\vect{\gamma}) \Big], \\
			&\vect{\hat{P}}_{t,t-j} \defeq \text{Cov} \, \Big[\vect{\Phi}_t, \vect{\Phi}_{t-j} \,|\, \mathscr{Y}(s), \vect{\hat{\vartheta}}_{s}^{k}(\vect{\gamma}) \Big],
		\end{align*}
		
		\noindent for any $k \geq 0$, $0 \leq j \leq t$ and $t \geq 0$. Let also $\vect{\hat{P}}_{t} \equiv \vect{\hat{P}}_{t,t}$.
	\end{definition}
	
	\begin{remark}
		These estimates are computed as in \cite{pellegrino2020selecting}.
	\end{remark}
	
	\noindent We also use the notation in \cref{def:observed_measurements} to further deal with missing observations.
	
	\begin{definition}[Observed measurements] \label{def:observed_measurements}
		Recall that $\mathscr{T} = \bigcup_{i=1}^N \mathscr{T}_i$ and let
		\begin{align*}
			&\mathscr{T}(s) \defeq \{t : t \in \mathscr{T}, \, 1 \leq t \leq s\},
		\end{align*}
		
		\noindent for $1 \leq s \leq T$. Let also
		\begin{align*}
			\mathscr{D}_{t} \defeq \{i : t \in \mathscr{T}_i, \, 1 \leq i \leq NK\},
		\end{align*}
		
		\noindent for $1 \leq t \leq T$. Finally, let
		\begin{align*}
			&\vect{Y}_{t}^{obs} \defeq \big( Y_{i,t} \big)_{i \in \mathscr{D}_t} \\
			&\vect{B}_{t}^{obs} \defeq \vect{A}_{t} \vect{B}
		\end{align*}
		
		\noindent be the $|\mathscr{D}_t| \times 1$ vector of observed measurements at time $t$ and the corresponding $|\mathscr{D}_t| \times q$ matrix of coefficients, for any $t \in \mathscr{T}$. Every $\vect{A}_t$ is indeed a selection matrix constituted by ones and zeros that permits to retrieve the appropriate rows of $\vect{B}$ for every $t \in \mathscr{T}$.
	\end{definition}

	\noindent Moreover, in order to simplify the notation, we let 
	\begin{align*}
			\mathcal{L}_e \left[\underline{\vect{\vartheta}} \,|\, \mathscr{Y}(s), \vect{\hat{\vartheta}}_{s}^{k}(\vect{\gamma}) \right] \equiv \expect \left [\mathcal{L}(\underline{\vect{\vartheta}} \,|\, \vect{Y}_{1:s}, \vect{\Phi}_{1:s}) \,|\, \mathscr{Y}(s), \vect{\hat{\vartheta}}_{s}^{k}(\vect{\gamma}) \right]
	\end{align*}
	
	\noindent for $1 \leq s \leq T$.
	
	It then follows directly from \citet[][Proposition 1]{pellegrino2021factoraugmented} that
	\begin{align*}
		\mathcal{L}_e \left[\underline{\vect{\vartheta}} \,|\, \mathscr{Y}(s), \vect{\hat{\vartheta}}_{s}^{k}(\vect{\gamma}) \right] \simeq 		&-\frac{1}{2}\ln|\underline{\vect{\Omega}_0}| - \frac{1}{2}\Tr\Big[\underline{\vect{\Omega}_0}^{-1} \big( \vect{\hat{\Phi}}_0 \vect{\hat{\Phi}}_0' + \vect{\hat{P}}_0 - \vect{\hat{\Phi}}_0 \underline{\vect{\mu}_0}' - \underline{\vect{\mu}_0}  \vect{\hat{\Phi}}_0' + \underline{\vect{\mu}_0} \, \underline{\vect{\mu}_0}' \big) \Big] \\
		&-\frac{s}{2}\ln|\underline{\vect{\Sigma}}| - \frac{1}{2} \Tr \Big[\underline{\vect{\Sigma}}^{-1} \big( \vect{\hat{E}}^{(1)}_s - \vect{\hat{E}}^{(2)}_s \underline{\vect{C}}_{\,*}' - \underline{\vect{C}}_{\,*} \vect{\hat{E}}^{(2)'}_s + \underline{\vect{C}}_{\,*} \vect{\hat{E}}^{(3)}_s \underline{\vect{C}}_{\,*}' \big) \Big] \\
		&-\frac{1}{2 \varepsilon} \Tr \left\{ \sum_{t \in \mathscr{T}(s)} \left[ \big( \vect{Y}_{t}^{obs} - \underline{\vect{B}}_{\,t}^{obs} \, \vect{\hat{\Phi}}_{t} \big) \big( \vect{Y}_{t}^{obs} - \underline{\vect{B}}_{\,t}^{obs} \, \vect{\hat{\Phi}}_{t} \big)' + \underline{\vect{B}}_{\,t}^{obs} \, \vect{\hat{P}}_{t} \, \underline{\vect{B}}_{\,t}^{{obs}'} \right] \right\},
	\end{align*}
	
	\noindent where
	\begin{alignat*}{3}
		&\vect{\hat{E}}^{(1)}_s &&\defeq \sum_{t=1}^s \expect\Big[\vect{\Phi}_{1:r, t} \vect{\Phi}_{1:r, t}' \,|\, \mathscr{Y}(s), \vect{\hat{\vartheta}}_{s}^{k}(\vect{\gamma}) \Big] &&= \sum_{t=1}^s \left(\vect{\hat{\Phi}}_t \vect{\hat{\Phi}}_t' + \vect{\hat{P}}_t\right)_{1:r, 1:r}, \\
		&\vect{\hat{E}}^{(2)}_s &&\defeq \sum_{t=1}^s \expect\Big[\vect{\Phi}_{1:r, t} \vect{\Phi}_{t-1}' \,|\, \mathscr{Y}(s), \vect{\hat{\vartheta}}_{s}^{k}(\vect{\gamma}) \Big] &&= \sum_{t=1}^s \left(\vect{\hat{\Phi}}_t \vect{\hat{\Phi}}_{t-1}' + \vect{\hat{P}}_{t, t-1}\right)_{1:r, 1:q}, \\
		&\vect{\hat{E}}^{(3)}_s &&\defeq \sum_{t=1}^s \expect\Big[\vect{\Phi}_{t-1} \vect{\Phi}_{t-1}' \,|\, \mathscr{Y}(s), \vect{\hat{\vartheta}}_{s}^{k}(\vect{\gamma}) \Big] &&= \sum_{t=1}^s \left(\vect{\hat{\Phi}}_{t-1} \vect{\hat{\Phi}}_{t-1}' + \vect{\hat{P}}_{t-1}\right),
	\end{alignat*}

	\noindent for $1 \leq s \leq T$. Furthermore, it follows from  \citet[][Lemma 1]{pellegrino2021factoraugmented} that
	\begin{align*}
		\expect \left[ \mathcal{P}(\underline{\vect{\vartheta}}, \vect{\gamma}) \,|\, \mathscr{Y}(s), \vect{\hat{\vartheta}}_{s}^{k}(\vect{\gamma}) \right] = \mathcal{P}(\underline{\vect{\vartheta}}, \vect{\gamma}).
	\end{align*}
	
	The CM-step conditionally maximises the expected penalised log-likelihood
	\begin{align}
		\mathcal{M}_e \left[\underline{\vect{\vartheta}}, \vect{\gamma} \,|\, \mathscr{Y}(s), \vect{\hat{\vartheta}}_{s}^{k}(\vect{\gamma}) \right] \defeq \mathcal{L}_e \left[\underline{\vect{\vartheta}} \,|\, \mathscr{Y}(s), \vect{\hat{\vartheta}}_{s}^{k}(\vect{\gamma}) \right] - \mathcal{P}(\underline{\vect{\vartheta}}, \vect{\gamma}) \label{eq:ecm_objective}
	\end{align}
	
	\noindent with respect to the free parameters.
	
	The CM-steps for all free parameters of the transition equation are reported in \citet[][Lemmas 2\,--\,4]{pellegrino2021factoraugmented}. For clarity, we recall just the lemmas' statements in \crefrange{lemma:ecm_init}{lemma:mstep_sigma} with the adequate minimal notational changes.
	
	% Lemma CM-step: init cond
	\begin{lemma} \label{lemma:ecm_init}
		The ECM estimator at a generic iteration $k+1 > 0$ for $\vect{\mu}_0$ is
		\begin{align*}
			&\vect{\hat{\mu}}_{0,s}^{k+1}(\vect{\gamma}) = \vect{\hat{\Phi}}_0
		\end{align*}
		
		\noindent and the estimator for $\vect{\Omega}_0$, denoted with $\vect{\hat{\Omega}}_{0,s}^{k+1}(\vect{\gamma})$, is a sparse covariance matrix whose entries allowed to differ from zero are
		\begin{align*}
			&\Big[ \vect{\hat{\Omega}}_{0,s}^{k+1}(\vect{\gamma}) \Big]_{i,j} = \Big[\vect{\hat{P}}_0\Big]_{i,j},
		\end{align*}
		
		\noindent and the coordinates $(i,j)$ are those described in  \cref{assumption:ssm_init_cond}.
	\end{lemma}
	
	% Lemma CM-step: C star
	\begin{lemma} \label{lemma:ecm_c_star}
		Note that $\vect{\Gamma}(\vect{\gamma}, 1) = \rho$ and let $\vect{\tilde{\Gamma}}(\vect{\gamma})$ be a $q \times q$ sparse matrix whose non-zero elements are
		\begin{align*}
		\vect{\tilde{\Gamma}}(\vect{\gamma}) \defeq \begin{pmatrix}
			\cdot & \cdot & \cdot & \cdot \\
			\cdot & \rho \, \vect{I}_{8+|\mathscr{G}|} & \cdot & \cdot \\
			\cdot & \cdot & \vect{\Gamma}(\vect{\gamma}, p) & \cdot \\
			\cdot& \cdot & \cdot & \cdot
		\end{pmatrix}.
	\end{align*}
		
		\noindent Moreover, let 
		\begin{align*}
			\mathscr{U}_{C} &\defeq \{1, \ldots, r\} \times \{1, \ldots, q\}, \\
			\mathscr{U}_{\pi} &\defeq \{(i,j) : i=j \text{ and } 7 + |\mathscr{G}| < i < r\} \cup \{(i,j) : i=r \text{ and } r \leq j < r+p\},
		\end{align*}
		
		\noindent wherein the latter can be partitioned as $\{\mathscr{C}(i,j), (i,j), \mathscr{C}''(i,j)\}$ for any $(i,j) \in \mathscr{U}_{\pi}$. Hence, the ECM estimator at a generic iteration $k+1 > 0$ for $\vect{C}$ is such that
		\begin{align*}
			\hat{C}_{i,j,s}^{k+1}(\vect{\gamma}) = \frac{\mathcal{S} \, \left[ \hat{\Sigma}_{i, i, s}^{k^{-1}}(\vect{\gamma}) \, \hat{E}^{(2)}_{i, j, s} - \sum_{\substack{(l_1, l_2) \in \mathscr{U}_{C} \\ (l_1, l_2) \neq (i,j)}} \hat{\Sigma}_{i, l_1, s}^{k^{-1}} (\vect{\gamma}) \, \hat{C}_{l_1,l_2, s}^{k+\indicator_{(l_1,l_2) \in \mathscr{C}(i,j)}}(\vect{\gamma})\, \hat{E}^{(3)}_{l_2, j, s}, \; \frac{\alpha}{2} \, \tilde{\Gamma}_{j,j}\,(\vect{\gamma}) \right]}{\hat{\Sigma}_{i,i,s}^{k^{-1}} (\vect{\gamma}) \, \hat{E}^{(3)}_{j,j,s} + (1-\alpha) \, \tilde{\Gamma}_{j,j}\,(\vect{\gamma})},
		\end{align*}
		
		\noindent for any $(i,j) \in \mathscr{U}_{\pi}$ and constant to the values in \cref{appendix:ecm:ssm} for the remaining entries, and with $\mathcal{S}$ being the soft-thresholding operator.
	\end{lemma}
	
	\begin{lemma} \label{lemma:mstep_sigma}
		The ECM estimator at a generic iteration $k+1 > 0$ for $\vect{\Sigma}$ is such that
		\begin{align*}
			&\hat{\Sigma}_{i,i,s}^{k+1}(\vect{\gamma}) = \frac{1}{s} \left[\vect{\hat{E}}^{(1)}_s - \vect{\hat{E}}^{(2)}_s \vect{\hat{C}}^{{k+1}'}_{s}(\vect{\gamma}) - \vect{\hat{C}}^{k+1}_{s}(\vect{\gamma}) \, \vect{\hat{E}}^{(2)'}_s + \vect{\hat{C}}^{k+1}_{s} (\vect{\gamma}) \, \vect{\hat{E}}^{(3)}_s \vect{\hat{C}}^{{k+1}'}_{s} (\vect{\gamma}) \right]_{i,i}
		\end{align*}
		
		\noindent for $i=1, \ldots, r$ and zero for the remaining entries.
	\end{lemma}
	
	The CM-step for the free parameters of the measurement equation requires an ad-hoc approach due to the implicit equality constraints for the households described in \cref{sec:methodology:framework} -- i.e., all households within a given group have the same factor loadings. While linear constraints have been handled before in EM-like algorithms for time-series models \citep[for instance, in order to apply mixed frequency aggregation constrains in now-casting problems as in][]{banbura2014maximum}, our problem is a bit more complicated. Indeed, it is not advised to estimate an unconstrained version of the model and apply the restrictions ex-post, since each consumer unit is observed for very short periods of time. Hence, we handle this CM-step as the constrained optimisation problem in \cref{proposition:ecm_loadings}.\footnote{We do not need to use Lagrangian multipliers, given that the constraints can be implemented by directly plugging them into the expected log-likelihood via $\vect{B}$ as described in \cref{appendix:ecm:ssm}.}
	
	\begin{proposition} \label{proposition:ecm_loadings}
		Let
		\begin{alignat*}{2}
			&\vect{\ddot{A}}_{t} &&\defeq \vect{A}_{t}' \vect{A}_{t}, \\[0.5em]
			&\vect{\hat{F}}_{t} &&\defeq \vect{\hat{\Phi}}_{t} \vect{\hat{\Phi}}_{t}' + \vect{\hat{P}}_{t},  \\[0.5em]
			&\vect{\hat{G}}_{s} &&\defeq \sum_{t \in \mathscr{T}(s)} \vect{A}_{t}' 	\vect{Y}_{t}^{obs} \,\vect{\hat{\Phi}}_{t}',
		\end{alignat*}
		
		\noindent and
		\begin{alignat*}{2}
			&\hat{\CyBe}_{i, j, s} &&\defeq \sum_{k=\omega_{i}^{\dag}}^{\omega_{i}^{\ddag}} \hat{G}_{k, j, s}, \\
			&\CyEl_{i, j, t} &&\defeq \sum_{k=\omega_{i}^{\dag}}^{\omega_{i}^{\ddag}} \ddot{A}_{k, j, t}, \\
			&\CySha_{i, j, t} &&\defeq \sum_{k=\omega_{i}^{\dag}}^{\omega_{i}^{\ddag}}  \sum_{l=\omega_{j}^{\dag}}^{\omega_{j}^{\ddag}} \ddot{A}_{k, l, t},
		\end{alignat*}
	
		\noindent where $\omega_{i}^{\dag} \defeq 9+\sum_{1 \leq j < i-7} \omega_{j}$ and $\omega_{i}^{\ddag} \defeq 8+\sum_{1 \leq j \leq i-7} \omega_{j}$, for $i=8, \ldots, 7+|\mathscr{G}|$. Let also
		\begin{align*}
			\mathscr{U}_{\Lambda} &\defeq \{1, \ldots, 7+|\mathscr{G}|\} \times \{0, \ldots, p-1\}.
		\end{align*}
		
		\noindent It follows that, when the penalty is not active and at a generic $k+1$ iteration of the ECM algorithm, the factor loadings
		\begin{align*}
			\hat{\Lambda}_{\,i,j+1}^{QMLE, \, k+1} &= \frac{1}{\sum_{t \in \mathscr{T}(s)} \hat{F}_{j+r,j+r,t} \,  \ddot{A}_{i+1, i+1, t}} \, \Bigg\{ \hat{G}_{i+1,j+r,s} - \sum_{t \in \mathscr{T}(s)} \bigg[\sum_{l_1=1}^{NK} \sum_{l_2=1}^{r-1} \hat{F}_{j+r,l_2,t} \, B_{\,l_1,l_2} \,  \ddot{A}_{i+1,l_1,t} \\[0.5em]
			&\quad \qquad + \hat{F}_{j+r,r,t} \,  \ddot{A}_{1,i+1,t} + \sum_{\substack{(l_1, l_2) \in \mathscr{U}_{\Lambda} \\ (l_1, l_2) \neq (i, j)}} \hat{F}_{j+r,l_2+r,t} \, \hat{\Lambda}^{\diamond}_{\,l_1, l_2+1} \, \big( \indicator_{l_1 \leq 7} \, \ddot{A}_{l_1+1, i+1, t} + \indicator_{l_1 > 7} \, \CyEl_{l_1,i+1,t} \big) \bigg] \Bigg\}
		\end{align*}
		
		\noindent for $i=1, \ldots, 7$, and
		\begin{align*}
			\hat{\Lambda}_{\,i,j+1}^{QMLE, \, k+1}  &= \frac{1}{\sum_{t \in \mathscr{T}(s)} \hat{F}_{j+r,j+r,t} \,  \CySha_{i,i,t}} \, \Bigg\{\hat{\CyBe}_{i, j+r, s} - \sum_{t \in \mathscr{T}(s)}  \bigg[\sum_{l_1=1}^{NK} \sum_{l_2=1}^{r-1} \hat{F}_{j+r,l_2,t} \, B_{\,l_1,l_2} \,  \CyEl_{i,l_1,t} \\ 
			&\quad \qquad + \hat{F}_{r,j+r,t} \,  \CyEl_{i,1,t} + \sum_{\substack{(l_1, l_2) \in \mathscr{U}_{\Lambda} \\ (l_1, l_2) \neq (i, j)}} \hat{F}_{j+r,l_2+r,t} \, \hat{\Lambda}^{\diamond}_{\,l_1,l_2+1} \,  \big( \indicator_{l_1 \leq 7} \, \CyEl_{i,l_1+1,t} + \indicator_{l_1 > 7} \, \CySha_{l_1,i,t} \big) \bigg] \Bigg\}
		\end{align*}
		
		\noindent for $i=8, \ldots, 7+|\mathscr{G}|$ where $\hat{\Lambda}_{l_{1}, l_{2}+1}^{\diamond}$ is the most up-to-date estimate for $\Lambda_{l_{1}, l_{2}+1}$ available when computing $\hat{\Lambda}^{k+1}_{i, j+1}$, for $j=0, \ldots, p-1$. More generally, it follows that, at a generic $k+1$ iteration of the ECM algorithm, the factor loadings
		\begin{align*}
			\hat{\Lambda}^{k+1}_{\,i,j+1}(\vect{\gamma}) = \begin{cases} 
				\frac{\mathcal{S} \left[ \hat{\Lambda}_{\,i,j+1}^{QMLE, \, k+1} \sum_{t \in \mathscr{T}(s)} \hat{F}_{j+r,j+r,t} \,  \ddot{A}_{i+1, i+1, t}  \, ,\; \frac{\varepsilon \alpha}{2}  \Gamma_{j+1, j+1} (\vect{\gamma}, p) \right]}{\varepsilon (1-\alpha) \Gamma_{j+1, j+1} (\vect{\gamma}, p) \;+\; \sum_{t \in \mathscr{T}(s)} \hat{F}_{j+r,j+r,t} \,  \ddot{A}_{i+1, i+1, t}}, & 1 \leq i \leq 7, \\[1em]
				\frac{\mathcal{S} \left[ \hat{\Lambda}_{\,i,j+1}^{QMLE, \, k+1} \sum_{t \in \mathscr{T}(s)} \hat{F}_{j+r,j+r,t} \,  \CySha_{i,i,t}  \, ,\; \frac{\varepsilon \alpha}{2}  \Gamma_{j+1, j+1} (\vect{\gamma}, p) \right]}{\varepsilon (1-\alpha) \Gamma_{j+1, j+1} (\vect{\gamma}, p) \;+\; \sum_{t \in \mathscr{T}(s)} \hat{F}_{j+r,j+r,t} \,  \CySha_{i,i,t}}, & 8 \leq i \leq 7+|\mathscr{G}|,
			\end{cases}
		\end{align*}
		
		\noindent for $j=0, \ldots, p-1$.
	\end{proposition}

	\begin{proofenv}
		We develop the proof in three steps. Step (i) derives the part of the expected log-likelihood that depends on the factor loadings. Step (ii) solves the maximisation problem assuming that the penalty is not active (i.e., $\rho=0$). This leads to a CM-step similar to the M-step usually employed in non-regularised EM-algorithms for dynamic factor models such as \cite{banbura2014maximum}. Step (iii) builds on that to write down the formula for the final estimator.
		
		(i) Note that, for the linearity of the trace,
		\begin{align*}
			&\Tr \left\{ \sum_{t \in \mathscr{T}(s)} \left[ \left( \vect{Y}_{t}^{obs} - \underline{\vect{B}}_{\,t}^{obs} \, \vect{\hat{\Phi}}_{t} \right) \left( \vect{Y}_{t}^{obs} - \underline{\vect{B}}_{\,t}^{obs} \, \vect{\hat{\Phi}}_{t} \right)' + \underline{\vect{B}}_{\,t}^{obs} \, \vect{\hat{P}}_{t} \, \underline{\vect{B}}_{\,t}^{{obs}'} \right] \right\} \\
			&\qquad = \sum_{t \in \mathscr{T}(s)} \Tr \left[ \left( \vect{Y}_{t}^{obs} - \underline{\vect{B}}_{\,t}^{obs} \, \vect{\hat{\Phi}}_{t} \right) \left( \vect{Y}_{t}^{obs} - \underline{\vect{B}}_{\,t}^{obs} \, \vect{\hat{\Phi}}_{t} \right)' + \underline{\vect{B}}_{\,t}^{obs} \, \vect{\hat{P}}_{t} \, \underline{\vect{B}}_{\,t}^{{obs}'} \right].
		\end{align*}
	
		\noindent The part of this trace that depends on the measurement coefficients is
		\begin{align*}
			&\sum_{t \in \mathscr{T}(s)} \Tr \left( \underline{\vect{B}}_{\,t}^{obs} \, \vect{\hat{F}}_{t} \, \underline{\vect{B}}_{\,t}^{{obs}'} - \underline{\vect{B}}_{\,t}^{obs} \, \vect{\hat{\Phi}}_{t} \, \vect{Y}_{t}^{{obs}'} - \vect{Y}_{t}^{obs} \, \vect{\hat{\Phi}}_{t}' \, \underline{\vect{B}}_{\,t}^{{obs}'} \right) \\
			&\qquad = \sum_{t \in \mathscr{T}(s)} \Tr \left( \vect{A}_{t} \underline{\vect{B}} \, \vect{\hat{F}}_{t} \, \underline{\vect{B}}' \vect{A}_{t}' - \vect{A}_{t} \underline{\vect{B}} \, \vect{\hat{\Phi}}_{t} \, \vect{Y}_{t}^{{obs}'} - \vect{Y}_{t}^{obs} \, \vect{\hat{\Phi}}_{t}' \, \underline{\vect{B}}' \vect{A}_{t}' \right) \\
			&\qquad = \sum_{t \in \mathscr{T}(s)} \Tr \left( \vect{A}_{t} \underline{\vect{B}} \, \vect{\hat{F}}_{t} \, \underline{\vect{B}}' \vect{A}_{t}' - \underline{\vect{B}} \, \vect{\hat{\Phi}}_{t} \, \vect{Y}_{t}^{{obs}'} \, \vect{A}_{t}  - \vect{A}_{t}' \, \vect{Y}_{t}^{obs} \, \vect{\hat{\Phi}}_{t}' \, \underline{\vect{B}}' \right) \\
			&\qquad = \sum_{t \in \mathscr{T}(s)} \Tr \left( \underline{\vect{B}} \, \vect{\hat{F}}_{t} \, \underline{\vect{B}}' \, \vect{\ddot{A}}_{t} \right) - \Tr \left(\underline{\vect{B}} \, \vect{\hat{G}}_{s}' \right)  - \Tr \left(\vect{\hat{G}}_{s} \, \underline{\vect{B}}' \right) \\
			&\qquad = \sum_{t \in \mathscr{T}(s)} \Tr \left( \underline{\vect{B}} \, \vect{\hat{F}}_{t} \, \underline{\vect{B}}' \, \vect{\ddot{A}}_{t} \right) - 2 \Tr \left(\underline{\vect{B}} \, \vect{\hat{G}}_{s}' \right).
		\end{align*}
		
		\noindent In order to write down the CM-step for the factor loadings, we develop these traces as functions of $\underline{\vect{\Lambda}}$.\footnote{Indeed, these are the only free parameters in $\underline{\vect{B}}$.} We start with the simpler trace. Under the identification restrictions and constraints  in \cref{sec:methodology:framework},
		\begin{align*}
			&\Tr \left(\underline{\vect{B}} \, \vect{\hat{G}}_{s}' \right) \\
			&\qquad= \sum_{i=1}^{NK} \sum_{j=1}^{q}  \underline{B}_{\,i,j} \, \hat{G}_{i,j,s} \\
			&\qquad= \sum_{i=1}^{NK} \sum_{j=1}^{r-1}  \underline{B}_{\,i,j} \, \hat{G}_{i,j,s} + \sum_{i=1}^{NK} \sum_{j=r}^{q}  \underline{B}_{\,i,j} \, \hat{G}_{i,j,s} \\
			&\qquad= \sum_{i=1}^{NK} \sum_{j=1}^{r-1}  \underline{B}_{\,i,j} \, \hat{G}_{i,j,s} + \sum_{i=1}^{8} \sum_{j=r}^{q}  \underline{B}_{\,i,j} \, \hat{G}_{i,j,s} + \sum_{i=9}^{NK} \sum_{j=r}^{q}  \underline{B}_{\,i,j} \, \hat{G}_{i,j,s} \\
			&\qquad= \sum_{i=1}^{NK} \sum_{j=1}^{r-1}  \underline{B}_{\,i,j} \, \hat{G}_{i,j,s} + \sum_{i=1}^{8} \sum_{j=r}^{r+p-1}  \underline{B}_{\,i,j} \, \hat{G}_{i,j,s} + \sum_{i=9}^{NK} \sum_{j=r}^{r+p-1}  \underline{B}_{\,i,j} \, \hat{G}_{i,j,s} \\
			&\qquad= \sum_{i=1}^{NK} \sum_{j=1}^{r-1}  \underline{B}_{\,i,j} \, \hat{G}_{i,j,s} + \sum_{i=1}^{8} \sum_{j=0}^{p-1}  \underline{B}_{\,i,j+r} \, \hat{G}_{i,j+r,s} + \sum_{i=9}^{NK} \sum_{j=0}^{p-1}  \underline{B}_{\,i,j+r} \, \hat{G}_{i,j+r,s} \\
			&\qquad= \sum_{i=1}^{NK} \sum_{j=1}^{r-1}  \underline{B}_{\,i,j} \, \hat{G}_{i,j,s} \, + \, \hat{G}_{1,r,s} + \sum_{i=1}^{7} \sum_{j=0}^{p-1}  \underline{\Lambda}_{\,i, j+1} \, \hat{G}_{i+1,j+r,s} + \sum_{i=8}^{7+|\mathscr{G}|} \sum_{j=0}^{p-1} \underline{\Lambda}_{\,i, j+1} \hat{\CyBe}_{i, j+r, s}.
		\end{align*}
		
		\noindent Thus,
		\begin{align} \label{eq:trace_bg_1}
			\Tr \left(\underline{\vect{B}} \, \vect{\hat{G}}_{s}' \right)  \propto \sum_{i=1}^{7} \sum_{j=0}^{p-1}  \underline{\Lambda}_{\,i, j+1} \, \hat{G}_{i+1,j+r,s} + \sum_{i=8}^{7+|\mathscr{G}|} \sum_{j=0}^{p-1} \underline{\Lambda}_{\,i, j+1} \hat{\CyBe}_{i, j+r, s}.
		\end{align}
		
		\noindent Next, we focus on the most complicated component of the expected log-likelihood. Since $\vect{\ddot{A}}_{t}$ and $\vect{\hat{F}}_{t}$ are symmetric,
		\begin{align*}
			&\Tr \left( \underline{\vect{B}} \, \vect{\hat{F}}_{t} \, \underline{\vect{B}}' \, \vect{\ddot{A}}_{t} \right) \\
			&\qquad= \sum_{i=1}^{NK} \sum_{j=1}^{q} \sum_{k=1}^{q} \sum_{l=1}^{NK} \underline{B}_{\,i,j} \, \hat{F}_{j,k,t} \, \underline{B}_{\,l,k} \,  \ddot{A}_{l,i,t} \\
			&\qquad= \sum_{i=1}^{NK} \sum_{j=1}^{r-1} \sum_{k=1}^{r-1} \sum_{l=1}^{NK} \underline{B}_{\,i,j} \, \hat{F}_{j,k,t} \, \underline{B}_{\,l,k} \,  \ddot{A}_{l,i,t} + \sum_{i=1}^{NK} \sum_{j=r}^{q} \sum_{k=1}^{r-1} \sum_{l=1}^{NK} \underline{B}_{\,i,j} \, \hat{F}_{j,k,t} \, \underline{B}_{\,l,k} \,  \ddot{A}_{l,i,t} \\
			&\qquad \qquad + \sum_{i=1}^{NK} \sum_{j=1}^{r-1} \sum_{k=r}^{q} \sum_{l=1}^{NK} \underline{B}_{\,i,j} \, \hat{F}_{j,k,t} \, \underline{B}_{\,l,k} \,  \ddot{A}_{l,i,t} + \sum_{i=1}^{NK} \sum_{j=r}^{q} \sum_{k=r}^{q} \sum_{l=1}^{NK} \underline{B}_{\,i,j} \, \hat{F}_{j,k,t} \, \underline{B}_{\,l,k} \,  \ddot{A}_{l,i,t} \\
			&\qquad= \sum_{i=1}^{NK} \sum_{j=1}^{r-1} \sum_{k=1}^{r-1} \sum_{l=1}^{NK} \underline{B}_{\,i,j} \, \hat{F}_{j,k,t} \, \underline{B}_{\,l,k} \,  \ddot{A}_{l,i,t} + 2\sum_{i=1}^{NK} \sum_{j=r}^{q} \sum_{k=1}^{r-1} \sum_{l=1}^{NK} \underline{B}_{\,i,j} \, \hat{F}_{j,k,t} \, \underline{B}_{\,l,k} \,  \ddot{A}_{l,i,t} \\
			&\qquad \qquad + \sum_{i=1}^{NK} \sum_{j=r}^{q} \sum_{k=r}^{q} \sum_{l=1}^{NK} \underline{B}_{\,i,j} \, \hat{F}_{j,k,t} \, \underline{B}_{\,l,k} \,  \ddot{A}_{l,i,t}.
		\end{align*}
		
		\noindent Thus, the only part of the latter trace that depends on the factor loadings is
		\begin{align} \label{eq:trace_bobn_1}
			\sum_{i=1}^{NK} \sum_{j=r}^{q} \sum_{k=r}^{q} \sum_{l=1}^{NK} \underline{B}_{\,i,j} \, \hat{F}_{j,k,t} \, \underline{B}_{\,l,k} \,  \ddot{A}_{l,i,t} + 2\sum_{i=1}^{NK} \sum_{j=r}^{q} \sum_{k=1}^{r-1} \sum_{l=1}^{NK} \underline{B}_{\,i,j} \, \hat{F}_{j,k,t} \, \underline{B}_{\,l,k} \,  \ddot{A}_{l,i,t}.
		\end{align} 
		
		\noindent The first term of \cref{eq:trace_bobn_1} is
		\begin{align*}
			&\sum_{i=1}^{NK} \sum_{j=r}^{q} \sum_{k=r}^{q} \sum_{l=1}^{NK} \underline{B}_{\,i,j} \, \hat{F}_{j,k,t} \, \underline{B}_{\,l,k} \,  \ddot{A}_{l,i,t} \\
			&\qquad= \sum_{i=1}^{8} \sum_{j=r}^{q} \sum_{k=r}^{q} \sum_{l=1}^{8} \underline{B}_{\,i,j} \, \hat{F}_{j,k,t} \, \underline{B}_{\,l,k} \,  \ddot{A}_{l,i,t} + \sum_{i=9}^{NK} \sum_{j=r}^{q} \sum_{k=r}^{q} \sum_{l=9}^{NK} \underline{B}_{\,i,j} \, \hat{F}_{j,k,t} \, \underline{B}_{\,l,k} \,  \ddot{A}_{l,i,t} \\
			&\qquad \qquad + \sum_{i=1}^{8} \sum_{j=r}^{q} \sum_{k=r}^{q} \sum_{l=9}^{NK} \underline{B}_{\,i,j} \, \hat{F}_{j,k,t} \, \underline{B}_{\,l,k} \,  \ddot{A}_{l,i,t} + \sum_{i=9}^{NK} \sum_{j=r}^{q} \sum_{k=r}^{q} \sum_{l=1}^{8} \underline{B}_{\,i,j} \, \hat{F}_{j,k,t} \, \underline{B}_{\,l,k} \,  \ddot{A}_{l,i,t} \\
			&\qquad= \sum_{i=1}^{8} \sum_{j=r}^{q} \sum_{k=r}^{q} \sum_{l=1}^{8} \underline{B}_{\,i,j} \, \hat{F}_{j,k,t} \, \underline{B}_{\,l,k} \,  \ddot{A}_{l,i,t} + \sum_{i=9}^{NK} \sum_{j=r}^{q} \sum_{k=r}^{q} \sum_{l=9}^{NK} \underline{B}_{\,i,j} \, \hat{F}_{j,k,t} \, \underline{B}_{\,l,k} \,  \ddot{A}_{l,i,t} \\
			&\qquad \qquad + 2\sum_{i=1}^{8} \sum_{j=r}^{q} \sum_{k=r}^{q} \sum_{l=9}^{NK} \underline{B}_{\,i,j} \, \hat{F}_{j,k,t} \, \underline{B}_{\,l,k} \,  \ddot{A}_{l,i,t}.
		\end{align*}
				
		\noindent Under the identification restrictions and constraints in \cref{sec:methodology:framework},
		\begin{align*}
			&\sum_{i=1}^{8} \sum_{j=r}^{q} \sum_{k=r}^{q} \sum_{l=1}^{8} \underline{B}_{\,i,j} \, \hat{F}_{j,k,t} \, \underline{B}_{\,l,k} \,  \ddot{A}_{l,i,t} \\
			&\qquad = \hat{F}_{r,r,t} \, \ddot{A}_{1,1,t} + 2 \sum_{i=2}^{8} \sum_{j=r}^{q} \underline{B}_{\,i,j} \, \hat{F}_{j,r,t} \,  \ddot{A}_{1,i,t} + \sum_{i=2}^{8} \sum_{j=r}^{q} \sum_{k=r}^{q} \sum_{l=2}^{8} \underline{B}_{\,i,j} \, \hat{F}_{j,k,t} \, \underline{B}_{\,l,k} \,  \ddot{A}_{l,i,t} \\
			&\qquad = \hat{F}_{r,r,t} \, \ddot{A}_{1,1,t} + 2 \sum_{i=2}^{8} \sum_{j=r}^{r+p-1} \underline{B}_{\,i,j} \, \hat{F}_{j,r,t} \,  \ddot{A}_{1,i,t} + \sum_{i=2}^{8} \sum_{j=r}^{r+p-1} \sum_{k=r}^{r+p-1} \sum_{l=2}^{8} \underline{B}_{\,i,j} \, \hat{F}_{j,k,t} \, \underline{B}_{\,l,k} \,  \ddot{A}_{l,i,t} \\
			&\qquad = \hat{F}_{r,r,t} \, \ddot{A}_{1,1,t} + 2 \sum_{i=2}^{8} \sum_{j=0}^{p-1} \underline{B}_{\,i,j+r} \, \hat{F}_{j+r,r,t} \,  \ddot{A}_{1,i,t} + \sum_{i=2}^{8} \sum_{j=0}^{p-1} \sum_{k=0}^{p-1} \sum_{l=2}^{8} \underline{B}_{\,i,j+r} \, \hat{F}_{j+r,k+r,t} \, \underline{B}_{\,l,k+r} \,  \ddot{A}_{l,i,t} \\
			&\qquad = \hat{F}_{r,r,t} \, \ddot{A}_{1,1,t} + 2 \sum_{i=1}^{7} \sum_{j=0}^{p-1} \underline{\Lambda}_{\,i,j+1} \, \hat{F}_{j+r,r,t} \,  \ddot{A}_{1,i+1,t} + \sum_{i=1}^{7} \sum_{j=0}^{p-1} \sum_{k=0}^{p-1} \sum_{l=1}^{7} \underline{\Lambda}_{\,i,j+1} \, \hat{F}_{j+r,k+r,t} \, \underline{\Lambda}_{\,l,k+1} \,  \ddot{A}_{l+1, i+1, t}.
		\end{align*}
			
		\noindent Also,
		\begin{align*}
			&\sum_{i=9}^{NK} \sum_{j=r}^{q} \sum_{k=r}^{q} \sum_{l=9}^{NK} \underline{B}_{\,i,j} \, \hat{F}_{j,k,t} \, \underline{B}_{\,l,k} \,  \ddot{A}_{l,i,t} \\
			&\qquad = \sum_{i=9}^{NK} \sum_{j=r}^{r+p-1} \sum_{k=r}^{r+p-1} \sum_{l=9}^{NK} \underline{B}_{\,i,j} \, \hat{F}_{j,k,t} \, \underline{B}_{\,l,k} \,  \ddot{A}_{l,i,t} \\
			&\qquad = \sum_{i=9}^{NK} \sum_{j=0}^{p-1} \sum_{k=0}^{p-1} \sum_{l=9}^{NK} \underline{B}_{\,i,j+r} \, \hat{F}_{j+r,k+r,t} \, \underline{B}_{\,l,k+r} \,  \ddot{A}_{l,i,t} \\
			&\qquad = \sum_{i=8}^{7+|\mathscr{G}|} \sum_{j=0}^{p-1} \sum_{k=0}^{p-1} \sum_{l=8}^{7+|\mathscr{G}|} \underline{\Lambda}_{\,i,j+1} \, \hat{F}_{j+r,k+r,t} \, \underline{\Lambda}_{\,l,k+1} \,  \CySha_{l,i,t}.
		\end{align*}
		
		\noindent Moreover,
		\begin{align*}
			&\sum_{i=1}^{8} \sum_{j=r}^{q} \sum_{k=r}^{q} \sum_{l=9}^{NK} \underline{B}_{\,i,j} \, \hat{F}_{j,k,t} \, \underline{B}_{\,l,k} \,  \ddot{A}_{l,i,t} \\
			&\qquad = \sum_{i=1}^{8} \sum_{j=r}^{r+p-1} \sum_{k=r}^{r+p-1} \sum_{l=9}^{NK} \underline{B}_{\,i,j} \, \hat{F}_{j,k,t} \, \underline{B}_{\,l,k} \,  \ddot{A}_{l,i,t} \\
			&\qquad = \sum_{i=1}^{8} \sum_{j=0}^{p-1} \sum_{k=0}^{p-1} \sum_{l=9}^{NK} \underline{B}_{\,i,j+r} \, \hat{F}_{j+r,k+r,t} \, \underline{B}_{\,l,k+r} \,  \ddot{A}_{l,i,t} \\
			&\qquad = \sum_{k=0}^{p-1} \sum_{l=9}^{NK} \hat{F}_{r,k+r,t} \, \underline{B}_{\,l,k+r} \,  \ddot{A}_{l,1,t} + \sum_{i=2}^{8} \sum_{j=0}^{p-1} \sum_{k=0}^{p-1} \sum_{l=9}^{NK} \underline{B}_{\,i,j+r} \, \hat{F}_{j+r,k+r,t} \, \underline{B}_{\,l,k+r} \,  \ddot{A}_{l,i,t} \\
			&\qquad = \sum_{i=9}^{NK} \sum_{j=0}^{p-1} \hat{F}_{r,j+r,t} \, \underline{B}_{\,i,j+r} \,  \ddot{A}_{i,1,t} + \sum_{i=2}^{8} \sum_{j=0}^{p-1} \sum_{k=0}^{p-1} \sum_{l=9}^{NK} \underline{B}_{\,i,j+r} \, \hat{F}_{j+r,k+r,t} \, \underline{B}_{\,l,k+r} \,  \ddot{A}_{l,i,t} \\
			&\qquad = \sum_{i=8}^{7+|\mathscr{G}|} \sum_{j=0}^{p-1} \hat{F}_{r,j+r,t} \, \underline{\Lambda}_{\,i,j+1} \,  \CyEl_{i,1,t} + \sum_{i=1}^{7} \sum_{j=0}^{p-1} \sum_{k=0}^{p-1} \sum_{l=8}^{7+|\mathscr{G}|} \underline{\Lambda}_{\,i,j+1} \, \hat{F}_{j+r,k+r,t} \, \underline{\Lambda}_{\,l,k+1} \,  \CyEl_{l,i+1,t}.
		\end{align*}
	
		\noindent Thus, the first term of  \cref{eq:trace_bobn_1} is proportional to
		\begin{align} \label{eq:trace_bobn_2}
			&2 \sum_{i=1}^{7} \sum_{j=0}^{p-1} \underline{\Lambda}_{\,i,j+1} \, \hat{F}_{j+r,r,t} \,  \ddot{A}_{1,i+1,t} + \sum_{i=1}^{7} \sum_{j=0}^{p-1} \sum_{k=0}^{p-1} \sum_{l=1}^{7} \underline{\Lambda}_{\,i,j+1} \, \hat{F}_{j+r,k+r,t} \, \underline{\Lambda}_{\,l,k+1} \,  \ddot{A}_{l+1, i+1, t} \\
			&\qquad +  \sum_{i=8}^{7+|\mathscr{G}|} \sum_{j=0}^{p-1} \sum_{k=0}^{p-1} \sum_{l=8}^{7+|\mathscr{G}|} \underline{\Lambda}_{\,i,j+1} \, \hat{F}_{j+r,k+r,t} \, \underline{\Lambda}_{\,l,k+1} \,  \CySha_{l,i,t} + 2 \sum_{i=8}^{7+|\mathscr{G}|} \sum_{j=0}^{p-1} \hat{F}_{r,j+r,t} \, \underline{\Lambda}_{\,i,j+1} \,  \CyEl_{i,1,t} \nonumber \\
			&\qquad \qquad + 2 \sum_{i=1}^{7} \sum_{j=0}^{p-1} \sum_{k=0}^{p-1} \sum_{l=8}^{7+|\mathscr{G}|} \underline{\Lambda}_{\,i,j+1} \, \hat{F}_{j+r,k+r,t} \, \underline{\Lambda}_{\,l,k+1} \,  \CyEl_{l,i+1,t}. \nonumber
		\end{align}
		
		\noindent Finally, the second term of \cref{eq:trace_bobn_1} is
		\begin{align*}
			&2\sum_{i=1}^{NK} \sum_{j=r}^{q} \sum_{k=1}^{r-1} \sum_{l=1}^{NK} \underline{B}_{\,i,j} \, \hat{F}_{j,k,t} \, \underline{B}_{\,l,k} \,  \ddot{A}_{l,i,t} \\
			&\qquad = 2\sum_{i=1}^{NK} \sum_{j=r}^{r+p-1} \sum_{k=1}^{r-1} \sum_{l=1}^{NK} \underline{B}_{\,i,j} \, \hat{F}_{j,k,t} \, \underline{B}_{\,l,k} \,  \ddot{A}_{l,i,t} \\
			&\qquad = 2\sum_{i=1}^{NK} \sum_{j=0}^{p-1} \sum_{k=1}^{r-1} \sum_{l=1}^{NK} \underline{B}_{\,i,j+r} \, \hat{F}_{j+r,k,t} \, \underline{B}_{\,l,k} \,  \ddot{A}_{l,i,t} \\
			&\qquad = 2\sum_{i=1}^{NK} \sum_{j=0}^{p-1} \sum_{k=1}^{r-1} \sum_{l=1}^{NK} \underline{B}_{\,i,j+r} \, \hat{F}_{j+r,k,t} \, \underline{B}_{\,l,k} \,  \ddot{A}_{i,l,t} \\
			&\qquad = 2\sum_{i=1}^{8} \sum_{j=0}^{p-1} \sum_{k=1}^{r-1} \sum_{l=1}^{NK} \underline{B}_{\,i,j+r} \, \hat{F}_{j+r,k,t} \, \underline{B}_{\,l,k} \,  \ddot{A}_{i,l,t} + 2\sum_{i=9}^{NK} \sum_{j=0}^{p-1} \sum_{k=1}^{r-1} \sum_{l=1}^{NK} \underline{B}_{\,i,j+r} \, \hat{F}_{j+r,k,t} \, \underline{B}_{\,l,k} \,  \ddot{A}_{i,l,t},
		\end{align*}
	
		\noindent where
		\begin{align*}
			&\sum_{i=1}^{8} \sum_{j=0}^{p-1} \sum_{k=1}^{r-1} \sum_{l=1}^{NK} \underline{B}_{\,i,j+r} \, \hat{F}_{j+r,k,t} \, \underline{B}_{\,l,k} \,  \ddot{A}_{i,l,t} \\
			&\qquad = \sum_{k=1}^{r-1} \sum_{l=1}^{NK} \hat{F}_{r,k,t} \, \underline{B}_{\,l,k} \,  \ddot{A}_{1,l,t} + \sum_{i=2}^{8} \sum_{j=0}^{p-1} \sum_{k=1}^{r-1} \sum_{l=1}^{NK} \underline{B}_{\,i,j+r} \, \hat{F}_{j+r,k,t} \, \underline{B}_{\,l,k} \,  \ddot{A}_{i,l,t} \\
			&\qquad = \sum_{k=1}^{r-1} \sum_{l=1}^{NK} \hat{F}_{r,k,t} \, \underline{B}_{\,l,k} \,  \ddot{A}_{1,l,t} + \sum_{i=1}^{7} \sum_{j=0}^{p-1} \sum_{k=1}^{r-1} \sum_{l=1}^{NK} \underline{\Lambda}_{\,i,j+1} \, \hat{F}_{j+r,k,t} \, \underline{B}_{\,l,k} \,  \ddot{A}_{i+1,l,t}
		\end{align*}
	
		\noindent and
		\begin{align*}
			&\sum_{i=9}^{NK} \sum_{j=0}^{p-1} \sum_{k=1}^{r-1} \sum_{l=1}^{NK} \underline{B}_{\,i,j+r} \, \hat{F}_{j+r,k,t} \, \underline{B}_{\,l,k} \,  \ddot{A}_{i,l,t} = \sum_{i=8}^{7+|\mathscr{G}|} \sum_{j=0}^{p-1} \sum_{k=1}^{r-1} \sum_{l=1}^{NK} \underline{\Lambda}_{\,i,j+1} \, \hat{F}_{j+r,k,t} \, \underline{B}_{\,l,k} \,  \CyEl_{i,l,t}.
		\end{align*}
		
		\noindent Hence, the second term of \cref{eq:trace_bobn_1} is proportional to
		\begin{align}  \label{eq:trace_bobn_3}
			2\sum_{i=1}^{7} \sum_{j=0}^{p-1} \sum_{k=1}^{r-1} \sum_{l=1}^{NK} \underline{\Lambda}_{\,i,j+1} \, \hat{F}_{j+r,k,t} \, \underline{B}_{\,l,k} \,  \ddot{A}_{i+1,l,t} + 2\sum_{i=8}^{7+|\mathscr{G}|} \sum_{j=0}^{p-1} \sum_{k=1}^{r-1} \sum_{l=1}^{NK} \underline{\Lambda}_{\,i,j+1} \, \hat{F}_{j+r,k,t} \, \underline{B}_{\,l,k} \,  \CyEl_{i,l,t}.
		\end{align}
		
		\noindent Combining \crefrange{eq:trace_bobn_1}{eq:trace_bobn_3}, it follows that
		\begin{align} \label{eq:trace_bobn_4}
			&\Tr \left( \underline{\vect{B}} \, \vect{\hat{F}}_{t} \, \underline{\vect{B}}' \, \vect{\ddot{A}}_{t} \right) \\
			&\qquad \propto  2 \sum_{i=1}^{7} \sum_{j=0}^{p-1} \underline{\Lambda}_{\,i,j+1} \, \hat{F}_{j+r,r,t} \,  \ddot{A}_{1,i+1,t} + \sum_{i=1}^{7} \sum_{j=0}^{p-1} \sum_{k=0}^{p-1} \sum_{l=1}^{7} \underline{\Lambda}_{\,i,j+1} \, \hat{F}_{j+r,k+r,t} \, \underline{\Lambda}_{\,l,k+1} \,  \ddot{A}_{l+1, i+1, t} \nonumber \\
			&\qquad \qquad + 2 \sum_{i=8}^{7+|\mathscr{G}|} \sum_{j=0}^{p-1} \hat{F}_{r,j+r,t} \, \underline{\Lambda}_{\,i,j+1} \,  \CyEl_{i,1,t} + \sum_{i=8}^{7+|\mathscr{G}|} \sum_{j=0}^{p-1} \sum_{k=0}^{p-1} \sum_{l=8}^{7+|\mathscr{G}|} \underline{\Lambda}_{\,i,j+1} \, \hat{F}_{j+r,k+r,t} \, \underline{\Lambda}_{\,l,k+1} \,  \CySha_{l,i,t} \nonumber \\
			&\qquad \qquad \qquad + 2\sum_{i=1}^{7} \sum_{j=0}^{p-1} \sum_{k=1}^{r-1} \sum_{l=1}^{NK} \underline{\Lambda}_{\,i,j+1} \, \hat{F}_{j+r,k,t} \, \underline{B}_{\,l,k} \,  \ddot{A}_{i+1,l,t} + 2\sum_{i=8}^{7+|\mathscr{G}|} \sum_{j=0}^{p-1} \sum_{k=1}^{r-1} \sum_{l=1}^{NK} \underline{\Lambda}_{\,i,j+1} \, \hat{F}_{j+r,k,t} \, \underline{B}_{\,l,k} \,  \CyEl_{i,l,t} \nonumber \\
			&\qquad \qquad \qquad \qquad + 2 \sum_{i=1}^{7} \sum_{j=0}^{p-1} \sum_{k=0}^{p-1} \sum_{l=8}^{7+|\mathscr{G}|} \underline{\Lambda}_{\,i,j+1} \, \hat{F}_{j+r,k+r,t} \, \underline{\Lambda}_{\,l,k+1} \,  \CyEl_{l,i+1,t} \nonumber.
		\end{align}
		
		\noindent Finally, it follows from \cref{eq:trace_bg_1} and \cref{eq:trace_bobn_4} that
		\begin{align}\label{eq:clean_e_step_loadings}
			&\sum_{t \in \mathscr{T}(s)} \Tr \left( \underline{\vect{B}} \, \vect{\hat{F}}_{t} \, \underline{\vect{B}}' \, \vect{\ddot{A}}_{t} \right) - 2 \Tr \left(\underline{\vect{B}} \, \vect{\hat{G}}_{s}' \right) \\
			&\qquad \propto \sum_{t \in \mathscr{T}(s)} \sum_{i=1}^{7} \sum_{j=0}^{p-1} \bigg( 2 \underline{\Lambda}_{\,i,j+1} \, \hat{F}_{j+r,r,t} \,  \ddot{A}_{1,i+1,t} + \sum_{k=0}^{p-1} \sum_{l=1}^{7} \underline{\Lambda}_{\,i,j+1} \, \hat{F}_{j+r,k+r,t} \, \underline{\Lambda}_{\,l,k+1} \,  \ddot{A}_{l+1, i+1, t} \nonumber \\
			&\qquad \qquad + 2 \sum_{k=1}^{r-1} \sum_{l=1}^{NK} \underline{\Lambda}_{\,i,j+1} \, \hat{F}_{j+r,k,t} \, \underline{B}_{\,l,k} \,  \ddot{A}_{i+1,l,t} \bigg) -2 \sum_{i=1}^{7} \sum_{j=0}^{p-1}  \underline{\Lambda}_{\,i, j+1} \, \hat{G}_{i+1,j+r,s} \nonumber \\
			&\qquad \qquad \qquad + \sum_{t \in \mathscr{T}(s)} \sum_{i=8}^{7+|\mathscr{G}|} \sum_{j=0}^{p-1} \bigg( 2 \hat{F}_{r,j+r,t} \, \underline{\Lambda}_{\,i,j+1} \,  \CyEl_{i,1,t} + \sum_{k=0}^{p-1} \sum_{l=8}^{7+|\mathscr{G}|} \underline{\Lambda}_{\,i,j+1} \, \hat{F}_{j+r,k+r,t} \, \underline{\Lambda}_{\,l,k+1} \,  \CySha_{l,i,t} \nonumber \\
			&\qquad \qquad \qquad \qquad + 2 \sum_{k=1}^{r-1} \sum_{l=1}^{NK} \underline{\Lambda}_{\,i,j+1} \, \hat{F}_{j+r,k,t} \, \underline{B}_{\,l,k} \,  \CyEl_{i,l,t} \bigg) -2  \sum_{i=8}^{7+|\mathscr{G}|} \sum_{j=0}^{p-1} \underline{\Lambda}_{\,i, j+1} \hat{\CyBe}_{i, j+r, s} \nonumber \\
			&\qquad \qquad \qquad \qquad \qquad + 2 \sum_{t \in \mathscr{T}(s)} \sum_{i=1}^{7} \sum_{j=0}^{p-1} \sum_{k=0}^{p-1} \sum_{l=8}^{7+|\mathscr{G}|} \underline{\Lambda}_{\,i,j+1} \, \hat{F}_{j+r,k+r,t} \, \underline{\Lambda}_{\,l,k+1} \,  \CyEl_{l,i+1,t} \nonumber.
		\end{align}
		
		\noindent We have rearranged the terms in \cref{eq:clean_e_step_loadings} so that the first two rows refer to the factor loadings of the macroeconomic indicators, the third and fourth row refer to the ones of the households and the last row to both of them.
		
		(ii) When the penalty is not active, the CM-step is computed from \cref{eq:clean_e_step_loadings} since
		\begin{align*}
			\frac{\partial \mathcal{M}_e \left[\underline{\vect{\vartheta}}, \vect{\gamma} \,|\, \mathscr{Y}(s), \vect{\hat{\vartheta}}_{s}^{k}(\vect{\gamma}) \right]}{\partial \underline{\vect{\Lambda}}} = -\frac{1}{2 \varepsilon} \frac{\partial \left[\sum_{t \in \mathscr{T}(s)} \Tr \left( \underline{\vect{B}} \, \vect{\hat{F}}_{t} \, \underline{\vect{B}}' \, \vect{\ddot{A}}_{t} \right) - 2 \Tr \left(\underline{\vect{B}} \, \vect{\hat{G}}_{s}' \right)\right]}{\partial \underline{\vect{\Lambda}}}.
		\end{align*}
		
		\noindent We structure the CM-step by following analogous steps to those in \cite{pellegrino2021factoraugmented}. Indeed, we estimate $\vect{\Lambda}$ one entry at the time, starting from the $\Lambda_{1, 1}$ and in column-major order. In other words, the derivative of \cref{eq:clean_e_step_loadings} with respect to $\Lambda_{i,j+1}$ is taken having fixed the other factors loadings to their latest estimate, for any $i=1, \ldots, 7+|\mathscr{G}|$ and $j=0, \ldots, p-1$. Formally, at a generic $k+1$ iteration of the ECM algorithm, this derivative is equal to
		\begin{align*}
			&\frac{\hat{G}_{i+1,j+r,s}}{\varepsilon} - \frac{1}{\varepsilon} \sum_{t \in \mathscr{T}(s)} \Big(\sum_{l_1=1}^{NK} \sum_{l_2=1}^{r-1} \hat{F}_{j+r,l_2,t} \, B_{\,l_1,l_2} \,  \ddot{A}_{i+1,l_1,t} + \hat{F}_{j+r,r,t} \,  \ddot{A}_{1,i+1,t} + \underline{\Lambda}_{\,i,j+1} \hat{F}_{j+r,j+r,t} \,  \ddot{A}_{i+1, i+1, t} \\
			&\qquad + \sum_{\substack{(l_1, l_2) \in \mathscr{U}_{\Lambda} \\ (l_1, l_2) \neq (i, j) \\ l_1 \leq 7}} \hat{F}_{j+r,l_2+r,t} \, \hat{\Lambda}^{\diamond}_{\,l_1, l_2+1} \,  \ddot{A}_{l_1+1, i+1, t} + \sum_{\substack{(l_1, l_2) \in \mathscr{U}_{\Lambda} \\ (l_1, l_2) \neq (i, j) \\ l_1 > 7}} \hat{F}_{j+r,l_2+r,t} \, \hat{\Lambda}^{\diamond}_{\,l_1,l_2+1} \,  \CyEl_{l_1,i+1,t} \Big)
		\end{align*}
		
		\noindent when computed with respect to any factor loading associated to the macroeconomic aggregates, and
		\begin{align*}
			&\frac{\hat{\CyBe}_{i, j+r, s}}{\varepsilon} - \frac{1}{\varepsilon} \sum_{t \in \mathscr{T}(s)} \Big(\sum_{l_1=1}^{NK} \sum_{l_2=1}^{r-1} \hat{F}_{j+r,l_2,t} \, B_{\,l_1,l_2} \,  \CyEl_{i,l_1,t} +  \hat{F}_{r,j+r,t} \,  \CyEl_{i,1,t} + \underline{\Lambda}_{\,i,j+1} \, \hat{F}_{j+r,j+r,t} \,  \CySha_{i,i,t} \\ 
			&\qquad + \sum_{\substack{(l_1, l_2) \in \mathscr{U}_{\Lambda} \\ (l_1, l_2) \neq (i, j) \\ l_1 \leq 7}} \hat{\Lambda}^{\diamond}_{\,l_1,l_2+1} \, \hat{F}_{l_2+r,j+r,t} \, \CyEl_{i,l_1+1,t} + \sum_{\substack{(l_1, l_2) \in \mathscr{U}_{\Lambda} \\ (l_1, l_2) \neq (i, j) \\ l_1 > 7}} \hat{F}_{j+r,l_2+r,t} \, \hat{\Lambda}^{\diamond}_{\,l_1,l_2+1} \,  \CySha_{l_1,i,t} \Big)
		\end{align*}
	
		\noindent when computed with respect to any factor loading associated to the households data. These derivatives can be equivalently written in the compact forms
		\begin{align*}
			&\frac{\hat{G}_{i+1,j+r,s}}{\varepsilon} - \frac{1}{\varepsilon} \sum_{t \in \mathscr{T}(s)} \Big[\sum_{l_1=1}^{NK} \sum_{l_2=1}^{r-1} \hat{F}_{j+r,l_2,t} \, B_{\,l_1,l_2} \,  \ddot{A}_{i+1,l_1,t} + \hat{F}_{j+r,r,t} \,  \ddot{A}_{1,i+1,t} + \underline{\Lambda}_{\,i,j+1} \hat{F}_{j+r,j+r,t} \,  \ddot{A}_{i+1, i+1, t} \\
			&\qquad + \sum_{\substack{(l_1, l_2) \in \mathscr{U}_{\Lambda} \\ (l_1, l_2) \neq (i, j)}} \hat{F}_{j+r,l_2+r,t} \, \hat{\Lambda}^{\diamond}_{\,l_1, l_2+1} \, \big( \indicator_{l_1 \leq 7} \, \ddot{A}_{l_1+1, i+1, t} + \indicator_{l_1 > 7} \, \CyEl_{l_1,i+1,t} \big) \Big] \nonumber
		\end{align*}
		
		\noindent and
		\begin{align*}
			&\frac{\hat{\CyBe}_{i, j+r, s}}{\varepsilon} - \frac{1}{\varepsilon} \sum_{t \in \mathscr{T}(s)} \Big[\sum_{l_1=1}^{NK} \sum_{l_2=1}^{r-1} \hat{F}_{j+r,l_2,t} \, B_{\,l_1,l_2} \,  \CyEl_{i,l_1,t} +  \hat{F}_{r,j+r,t} \,  \CyEl_{i,1,t} + \underline{\Lambda}_{\,i,j+1} \, \hat{F}_{j+r,j+r,t} \,  \CySha_{i,i,t} \\ 
			&\qquad + \sum_{\substack{(l_1, l_2) \in \mathscr{U}_{\Lambda} \\ (l_1, l_2) \neq (i, j)}} \hat{F}_{j+r,l_2+r,t} \, \hat{\Lambda}^{\diamond}_{\,l_1,l_2+1} \,  \big( \indicator_{l_1 \leq 7} \, \CyEl_{i,l_1+1,t} + \indicator_{l_1 > 7} \, \CySha_{l_1,i,t} \big) \Big] \nonumber
		\end{align*}
		
		\noindent respectively. It follows that, when the penalty is not active and at a generic $k+1$ iteration of the ECM algorithm,
		\begin{align*}
			\hat{\Lambda}_{\,i,j+1}^{QMLE, \, k+1} &= \frac{1}{\sum_{t \in \mathscr{T}(s)} \hat{F}_{j+r,j+r,t} \,  \ddot{A}_{i+1, i+1, t}} \, \Bigg\{ \hat{G}_{i+1,j+r,s} - \sum_{t \in \mathscr{T}(s)} \bigg[\sum_{l_1=1}^{NK} \sum_{l_2=1}^{r-1} \hat{F}_{j+r,l_2,t} \, B_{\,l_1,l_2} \,  \ddot{A}_{i+1,l_1,t} \\[0.5em]
			&\quad \qquad + \hat{F}_{j+r,r,t} \,  \ddot{A}_{1,i+1,t} + \sum_{\substack{(l_1, l_2) \in \mathscr{U}_{\Lambda} \\ (l_1, l_2) \neq (i, j)}} \hat{F}_{j+r,l_2+r,t} \, \hat{\Lambda}^{\diamond}_{\,l_1, l_2+1} \, \big( \indicator_{l_1 \leq 7} \, \ddot{A}_{l_1+1, i+1, t} + \indicator_{l_1 > 7} \, \CyEl_{l_1,i+1,t} \big) \bigg] \Bigg\}
		\end{align*}
		
		\noindent for $i=1, \ldots, 7$, and
		\begin{align*}
			\hat{\Lambda}_{\,i,j+1}^{QMLE, \, k+1}  &= \frac{1}{\sum_{t \in \mathscr{T}(s)} \hat{F}_{j+r,j+r,t} \,  \CySha_{i,i,t}} \, \Bigg\{\hat{\CyBe}_{i, j+r, s} - \sum_{t \in \mathscr{T}(s)}  \bigg[\sum_{l_1=1}^{NK} \sum_{l_2=1}^{r-1} \hat{F}_{j+r,l_2,t} \, B_{\,l_1,l_2} \,  \CyEl_{i,l_1,t} \\ 
			&\quad \qquad + \hat{F}_{r,j+r,t} \,  \CyEl_{i,1,t} + \sum_{\substack{(l_1, l_2) \in \mathscr{U}_{\Lambda} \\ (l_1, l_2) \neq (i, j)}} \hat{F}_{j+r,l_2+r,t} \, \hat{\Lambda}^{\diamond}_{\,l_1,l_2+1} \,  \big( \indicator_{l_1 \leq 7} \, \CyEl_{i,l_1+1,t} + \indicator_{l_1 > 7} \, \CySha_{l_1,i,t} \big) \bigg] \Bigg\}
		\end{align*}
		
		\noindent for $i=8, \ldots, 7+|\mathscr{G}|$.
		
		(iii) It follows directly from the results in step (i) and step (ii), and the proof of \citet[][Lemma 5]{pellegrino2021factoraugmented} that, at a generic $k+1$ iteration of the ECM algorithm,
		\begin{align*}
			\hat{\Lambda}^{k+1}_{\,i,j+1}(\vect{\gamma}) = \begin{cases} 
				\frac{\mathcal{S} \left[ \hat{\Lambda}_{\,i,j+1}^{QMLE, \, k+1} \sum_{t \in \mathscr{T}(s)} \hat{F}_{j+r,j+r,t} \,  \ddot{A}_{i+1, i+1, t}  \, ,\; \frac{\varepsilon \alpha}{2}  \Gamma_{j+1, j+1} (\vect{\gamma}, p) \right]}{\varepsilon (1-\alpha) \Gamma_{j+1, j+1} (\vect{\gamma}, p) \;+\; \sum_{t \in \mathscr{T}(s)} \hat{F}_{j+r,j+r,t} \,  \ddot{A}_{i+1, i+1, t}}, & \text{if } 1 \leq i \leq 7, \\[1em]
				\frac{\mathcal{S} \left[ \hat{\Lambda}_{\,i,j+1}^{QMLE, \, k+1} \sum_{t \in \mathscr{T}(s)} \hat{F}_{j+r,j+r,t} \,  \CySha_{i,i,t}  \, ,\; \frac{\varepsilon \alpha}{2}  \Gamma_{j+1, j+1} (\vect{\gamma}, p) \right]}{\varepsilon (1-\alpha) \Gamma_{j+1, j+1} (\vect{\gamma}, p) \;+\; \sum_{t \in \mathscr{T}(s)} \hat{F}_{j+r,j+r,t} \,  \CySha_{i,i,t}}, & \text{if } 8 \leq i \leq 7+|\mathscr{G}|,
			\end{cases}
		\end{align*}
		
		\noindent for $j=0, \ldots, p-1$.
	\end{proofenv}
	
	\subsection{Initialisation} \label{appendix:ecm:init}
	First, we compute group averages for the microeconomic data. Then we apply the procedure described in \citet[][section A.3.]{pellegrino2021factoraugmented} on both the macroeconomic indices and group averages. For simplicity, we set $\lambda = 2.573$, $\alpha=0.667$ and $\beta=1.326$. These are the optimal values in \cite{pellegrino2021factoraugmented} converted for quarterly frequency data.
	
	\subsection{Enforcing causality during the estimation} \label{appendix:ecm:causality}
	We enforce causality during the estimation following \citet[][section A.4.]{pellegrino2021factoraugmented}.

	\clearpage
	\section{Additional charts} \label{appendix:charts}
	
\end{subappendices}

\end{document}